\documentclass[acmsmall]{acmart}

\AtBeginDocument{%
  }

\setcopyright{acmlicensed}
\acmJournal{TOIS}
\acmYear{2024} \acmVolume{1} \acmNumber{1} \acmArticle{1} \acmMonth{1}\acmDOI{10.1145/3643131}

\usepackage{subfig}
\usepackage{amsthm}
\usepackage{bm}
\usepackage{algorithm}
\usepackage{algorithmic}
\usepackage{amsfonts}

\begin{document}

\title{Can Perturbations Help Reduce Investment Risks? Risk-Aware Stock Recommendation via Split Variational Adversarial Training}

\author{Jiezhu Cheng}
\orcid{0000-0002-1755-6828}
\affiliation{%
  \institution{School of Computer Science and Engineering, Sun Yat-sen University}
  \streetaddress{No. 132, East of Outer Ring Road, Guangzhou Higher Education Mega Center, Panyu District}
  \city{Guangzhou}
  \state{Guangdong Province}
  \country{China}
  \postcode{510006}
}
\email{chengjzh@mail2.sysu.edu.cn}

\author{Kaizhu Huang}
\orcid{0000-0002-3034-9639}
\authornote{Corresponding author}
\affiliation{%
  \institution{Data Science Research Center, Duke Kunshan University}
  \streetaddress{No. 8 Duke Avenue, Kunshan}
  \city{Suzhou}
  \state{Jiangsu Province}
  \country{China}
  \postcode{215316}
}
\email{kaizhu.huang@dukekunshan.edu.cn}

\author{Zibin Zheng}
\orcid{0000-0002-7878-4330}
\affiliation{%
 \institution{School of Software Engineering, Sun Yat-sen University}
 \streetaddress{Tang Jia Wan}
 \city{Zhuhai}
 \state{Guangdong Province}
 \country{China}
 \postcode{519082}
}
\email{zhzibin@mail.sysu.edu.cn}

\renewcommand{\shortauthors}{Cheng, et al.}

\begin{abstract}
In the stock market, a successful investment requires a good balance between profits and risks. Based on the \emph{learning to rank} paradigm, stock recommendation has been widely studied in quantitative finance to recommend stocks with higher return ratios for investors. Despite the efforts to make profits, many existing recommendation approaches still have some limitations in risk control, which may lead to intolerable paper losses in practical stock investing. To effectively reduce risks, we draw inspiration from adversarial learning and propose a novel \emph{Split Variational Adversarial Training} (SVAT) method for risk-aware stock recommendation. Essentially, SVAT encourages the stock model to be sensitive to adversarial perturbations of risky stock examples and enhances the model's risk awareness by learning from perturbations. To generate representative adversarial examples as risk indicators, we devise a variational perturbation generator to model diverse risk factors. Particularly, the variational architecture enables our method to provide a rough risk quantification for investors, showing an additional advantage of interpretability. Experiments on several real-world stock market datasets demonstrate the superiority of our SVAT method. By lowering the volatility of the stock recommendation model, SVAT effectively reduces investment risks and outperforms state-of-the-art baselines by more than $30\%$ in terms of risk-adjusted profits. All the experimental data and source code are available at \url{https://drive.google.com/drive/folders/14AdM7WENEvIp5x5bV3zV_i4Aev21C9g6?usp=sharing}.
\end{abstract}

\ccsdesc[500]{Information systems~Data mining}
\ccsdesc[500]{Computing methodologies~Learning to rank}
\ccsdesc[500]{Computing methodologies~Adversarial learning}
\ccsdesc[500]{Applied computing~Economics}
\ccsdesc[300]{Computing methodologies~Machine learning}

\keywords{Stock recommendation, risk control, variational autoencoder}


\maketitle

\section{Introduction}
\label{sec:intro}

The stock market, one of the largest financial markets in the world, has been an attractive platform allowing millions of investors to manage their assets for wealth growth. However, its highly volatile nature presents not only opportunities for profits, but also risks of losses~\cite{stock_risk}. In order to achieve a good balance between profits and risks, stock investors have been striving for methods that can accurately predict the future trend of the stock market~\cite{stock_survey}. Unfortunately, stock prediction is extremely challenging due to the highly stochastic and non-stationary nature of stock prices. Under such circumstances, more and more researchers have opted for advanced machine learning methods to study stock movements and make profitable predictions~\cite{stock_survey_02}.

\begin{figure}[t]
  \centering
  \includegraphics[width=0.65\linewidth]{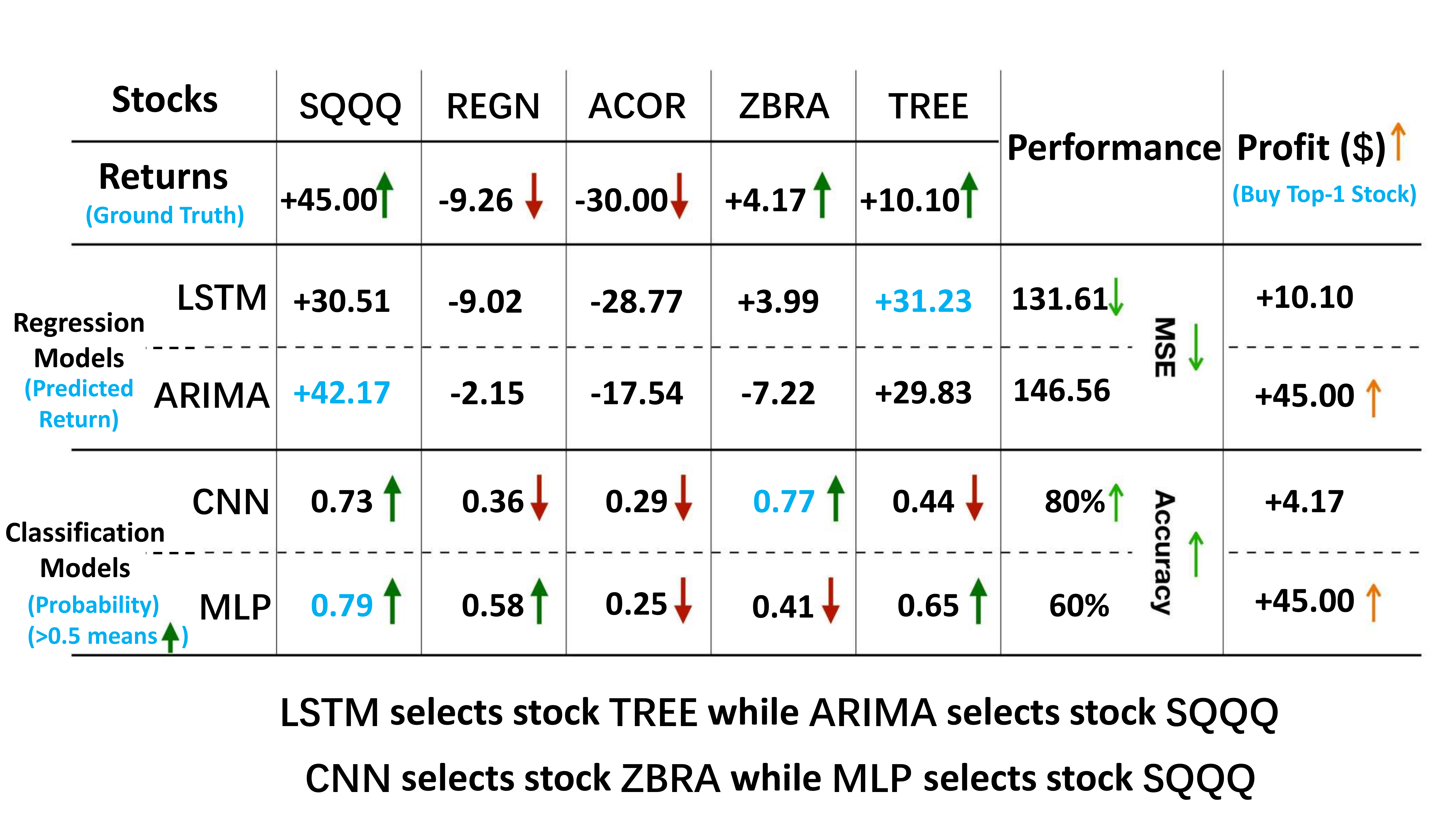}
  \caption{An example showing that for regression and classification methods, accurate stock prediction models LSTM (MSE $\downarrow$) and CNN (Acc. $\uparrow$) may earn less profit than inaccurate models ARIMA (MSE $\uparrow$) and MLP (Acc. $\downarrow$). The returns of the five stocks SQQQ, REGN, ACOR, ZBRA, and TREE are selected from the NASDAQ stock market at 02/23/2017.}
  \label{fig:stock_example}
  \Description{The stock example of abnormal results}
\end{figure}

Modern stock prediction solutions mainly fall into three categories, namely \emph{regression}, \emph{classification}, and \emph{recommendation} methods~\cite{stock_survey_02}. Regression methods formulate stock prediction as a pure time series forecasting problem and predict the future stock prices/returns by learning from historical stock time series data~\cite{text_stock_pred,stock_TS,tensor_stock_pred,DARNN,MLCNN,stockGAN_01,LogTrans}. On the other hand, classification methods treat stock prediction as a binary up/down classification problem and develop accurate classifiers to perform stock movement prediction~\cite{stocknet,adv_LSTM,Zibin_02}. Nevertheless, general regression and classification methods have a significant drawback that they are not directly optimized towards the target of investment (i.e., profit maximization)~\cite{RSR,STHAN-SR}, which may lead to abnormal results such that \emph{accurate prediction models earn less profit than inaccurate models}. Figure~\ref{fig:stock_example} shows an example of how the problem occurs. To overcome this drawback, some researchers have proposed to employ reinforcement learning methods~\cite{RL_trading_01,RL_trading_02,RL_trading_03} to improve model profits by capturing trading signals in a dynamic prediction. Other researchers have developed recommendation methods to rank stocks with return ratios based on the comparison among multiple stocks~\cite{TRAN}. In this case, models are trained to select top-$k$ stocks with maximum expected profits so as to ensure their consistency with the investment target. Accordingly, various stock recommendation models have been proposed and shown a promising prospect in the stock prediction domain~\cite{RSR,TRAN,STHAN-SR,ALSP_TF}.

\begin{figure}[t]
\centering

\subfloat[]{ 
  \begin{minipage}[t]{0.4\columnwidth}
  \centering
  \includegraphics[width=0.95\columnwidth]{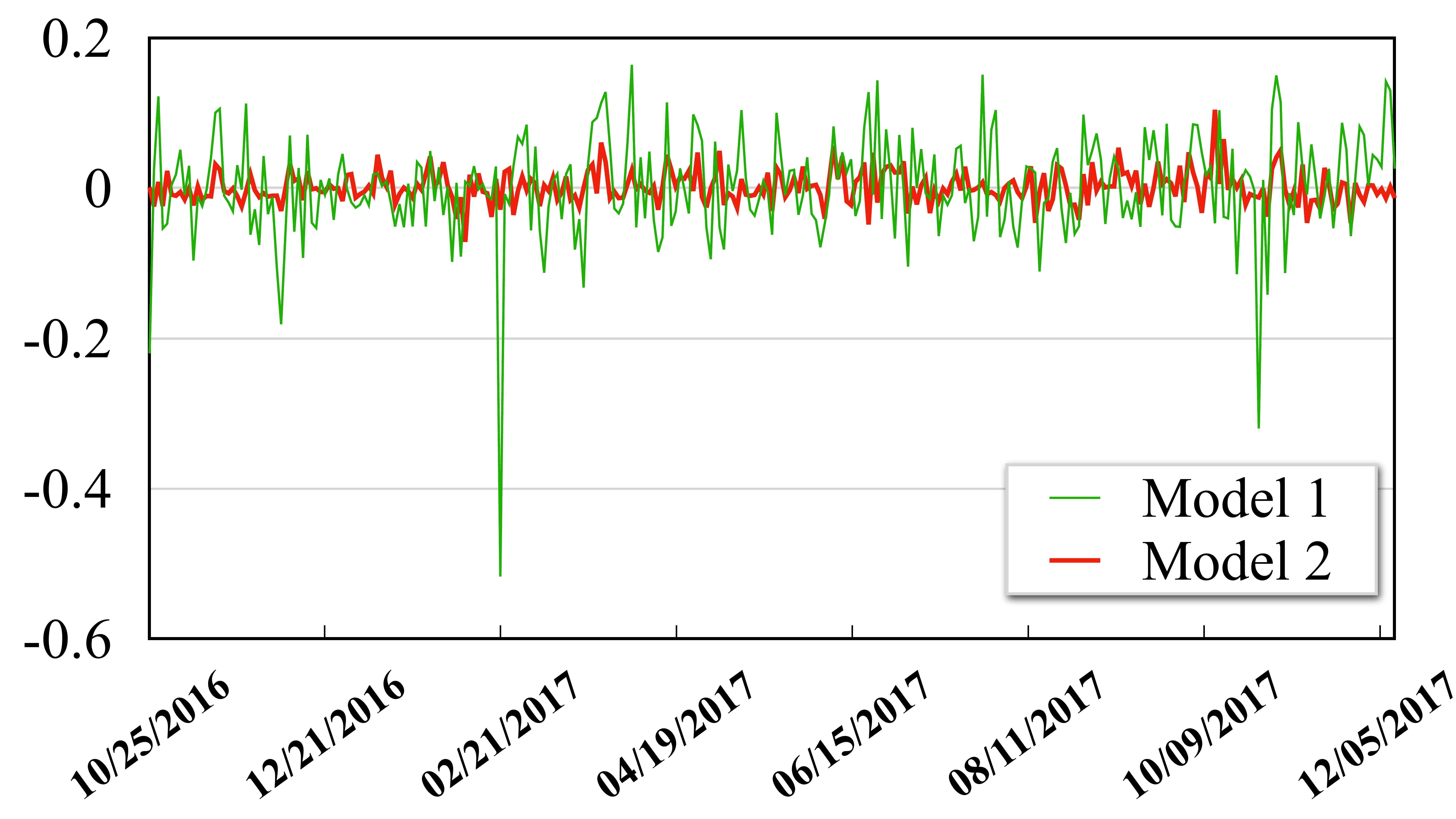}
  \label{fig:return_risk}
  \end{minipage}
}
\subfloat[]{
\raisebox{0.18\height}{
  \begin{minipage}[t]{0.5\columnwidth}
  \centering
  \includegraphics[width=0.88\columnwidth]{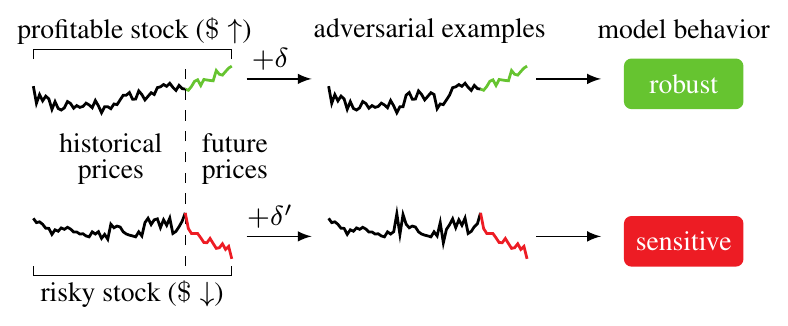}
  \label{fig:split_AT}
  \end{minipage}
}
}
\caption{(a) Daily returns of two stock recommendation models. Both models achieve similar total profits ($35.7\%$ for Model 1 and $35.2\%$ for Model 2) but with different volatilities. (b) Illustration of the split adversarial training. $\delta$, $\delta'$ are perturbations and $\$$ denotes the stock return. Best viewed in color.}
\Description{Motivation of the split AT design}
\label{fig:SAT_motivation}
\end{figure}

Despite the efforts to maximize profits for investors, many existing stock recommendation methods still have some limitations in \emph{risk control}. Most of them mainly focus on developing powerful learning models to improve the investment profit, while ignoring effective risk modeling. Such a deficiency may limit their effectiveness in practical stock investing and cause painful consequences. For example, Figure~\ref{fig:return_risk} presents daily returns of two stock recommendation models backtested in the NASDAQ stock market from 10/25/2016 to 12/11/2017. Although both models attain nearly the same amount of profit (i.e., the sum of all daily returns), Model~1 is more volatile than Model~2 and suffers from a higher risk of potential losses. When employing Model~1 for stock trading, even if the final profit ($35.7\%$) is considerable, the huge paper loss of $-51.7\%$ on 02/21/2017 can be intolerable to some investors and force them to stop investing halfway to prevent bankruptcy. In other words, the high volatility (risk) of Model~1 is prone to ``kill the investor before the dawn". To avoid such a disaster, it is imperative to reduce risks in addition to profit maximization when performing stock recommendation.

In this paper, we explore the possibility of leveraging adversarial perturbations~\cite{ADV_01} to reduce stock recommendation risks. This motivation leads to \underline{S}plit \underline{V}ariational \underline{A}dversarial \underline{T}raining (SVAT), a novel adversarial training (AT) framework for risk-aware stock recommendation. The first innovation of our method is the \emph{split} AT design. Without external information such as financial reports, the risk of potential losses mainly comes from adverse movements of historical stock prices, which are difficult to identify from the stochastic price series. To address the challenge, we propose to capture stock risks through the model's sensitivity to adversarial examples (AEs)~\cite{ADV_01}. As depicted in Figure~\ref{fig:split_AT}, for each stock example, we can generate AEs by adding small perturbations on their input features. Unlike conventional AT methods~\cite{ADV_02} which encourage the model to be robust to all AEs, we \emph{split the AT process} to make the model robust to AEs of \emph{profitable} stocks but sensitive to AEs of \emph{risky} stocks. In this way, the recommendation model can recognize risks by learning from different perturbations, improving the capability of risk control.

Another challenge is how to craft representative AEs as risk indicators. Stock prices can be affected by multiple risk factors such as company performance and macroeconomics, which cannot be effectively modeled by traditional gradient-based AT methods. Hence, motivated by Variational Autoencoders~\cite{VAE} and Adversarial Distributional Training~\cite{ADT}, we devise a variational perturbation generator (VPG) to learn an adversarial distribution driven by multiple latent risk factors, from which diverse perturbations can be generated for comprehensive risk modeling. Furthermore, in the testing environment, we could roughly quantify the risk of each stock example by generating its AEs from VPG and computing the entropy using different adversarial outputs. The risk quantification shows an additional advantage of interpretability provided by our method.

In this work, we employ STHAN-SR~\cite{STHAN-SR} as the backbone recommendation model and combine it with SVAT to achieve the state-of-the-art (SOTA) results. Our contributions are of four-folds:
\begin{itemize}
    \item We investigate the limitation of risk modeling negligence and highlight the necessity of reducing stock recommendation risks.
    \item We propose a novel Split Variational Adversarial Training method to enhance the risk sensitivity of the stock recommendation model, with the additional benefit of providing a rough risk quantification for investors.
    \item To the best of our knowledge, this is the first work to engage adversarial training for risk modeling in stock recommendation, showing a new possibility for reducing investment risks with adversarial perturbations.
    \item We conduct extensive experiments on three real-world datasets and show advantages of our model against state-of-the-art baselines, demonstrating the effectiveness and practicality of the SVAT method.
\end{itemize}

The remainder of this paper is organized as follows. Section~\ref{sec:preliminary} introduces the preliminary knowledge about stock recommendation and adversarial training, which forms the building blocks of our method. Section~\ref{sec:method} presents our proposed SVAT method. Section~\ref{sec:experiments} describes the extensive experiments we conduct. Finally, we review related work in Section~\ref{sec:related_work} and draw a conclusion in Section~\ref{sec:conclusion}.

\section{Preliminary}
\label{sec:preliminary}

\subsection{Problem Formulation}
\label{subsec:problem}

We focus on the task of stock recommendation under the learning to rank paradigm. Given a set of $N$ stocks $\mathcal{S} = \{s_1, s_2, ..., s_N\}$, for each stock $s_i \in \mathcal{S}$ on trading day $t$, there is an associated close price $p_{i,t}$ and a 1-day return ratio computed as $r_{i,t} = \frac{p_{i,t} - p_{i,t-1}}{p_{i,t-1}}$. According to the value of each $r_{i,t}$, we can determine a ranking list of all stocks sorted by their ranking scores $\mathbf{y}_t = \{y_{1,t} > y_{2,t} > ... > y_{N,t}\}$, where $y_{i,t} > y_{j,t}$ if and only if $r_{i,t} > r_{j,t}$ for any two stocks $s_i, s_j \in \mathcal{S}$. Therefore, stocks with higher ranking scores indicate higher investment revenue on trading day $t$. Formally, the goal of stock recommendation is to predict ranking scores $\hat{\mathbf{y}}_t$ given historical sequential data $\mathbf{X} = [X_{t-T}, X_{t-T+2}, ..., X_{t-1}]$:
\begin{equation}
\hat{\mathbf{y}}_t = f(\mathbf{X}; \Theta), \label{equ:stock_rank}
\end{equation}
where $X_{\tau} \in \mathbb{R}^{N \times d}$ represents the input features of all stocks on trading day $\tau$ ($\tau < t$), $d$ is the feature dimension and $T$ is the length of the lookback window. $f$ is the ranking function with parameters $\Theta$ to be learned. Following the previous work~\cite{RSR,STHAN-SR,ALSP_TF}, the loss function for optimizing $\Theta$ is the combination of a pointwise regression loss and a pairwise ranking loss:
\begin{equation}
\mathcal{L} = \sum_{i=1}^{N}(\hat{y}_{i,t} - y_{i,t})^2 + \alpha \sum_{i=1}^N \sum_{j=1}^N \max \{0, -(\hat{y}_{i,t} - \hat{y}_{j,t})(y_{i,t} - y_{j,t})\}, \label{equ:origin_loss}
\end{equation}
where $\alpha$ is a hyperparameter to balance the two loss terms. If not otherwise specified, we usually set the ground-truth ranking score $y_{i,t} = r_{i,t}$. After obtaining the prediction $\hat{\mathbf{y}}_t$, we can select the top-$k$ stocks from the ranking list for trading.

\subsection{Adversarial Training Definition}

Traditional adversarial training aims to improve the adversarial robustness of DNN classifiers by adding perturbations on sample features. Given a dataset $\mathcal{D} = \{(\mathbf{x}_i, y_i)\}_{i=1}^M$ of $M$ training examples with $\mathbf{x}_i \in \mathbb{R}^d$ (denoting the input features) and $y_i$ (denoting the ground-truth label), AT can be formulated as the following minimax optimization problem~\cite{PGD}:
\begin{equation}
\min_{\Theta} \frac{1}{M} \sum_{i=1}^M \max_{\bm{\delta}_i \in \Delta} \mathcal{L}(f(\mathbf{x}_i + \bm{\delta}_i; \Theta), y_i), \nonumber
\end{equation}
where $f$ is the DNN model with parameters $\Theta$, $\mathcal{L}$ is a loss function, $\Delta = \{\bm{\delta}: ||\bm{\delta}||_p \le \epsilon\}$ is a perturbation set with $\epsilon > 0$ and $||\cdot||_p$ denotes the $p$-norm of a vector. To enrich the diversity of perturbations, Adversarial Distributional Training (ADT)~\cite{ADT} is further proposed to learn a perturbation distribution $p(\bm{\delta}_i|\mathbf{x}_i)$ by minimizing
\begin{equation}
\min_{\Theta} \frac{1}{M} \sum_{i=1}^M \max_{p(\bm{\delta}_i|\mathbf{x}_i) \in \mathcal{P}} \mathbb{E}_{p(\bm{\delta}_i|\mathbf{x}_i)} [\mathcal{L}(f(\mathbf{x}_i + \bm{\delta}_i; \Theta), y_i)], \nonumber
\end{equation}
where $\mathcal{P} = \{p: \text{supp}(p) \subseteq \Delta \}$ is a set of distributions with support contained in $\Delta$. ADT solves the above minimax problem by simultaneously optimizing $p(\bm{\delta}_i|\mathbf{x}_i)$ and $\Theta$ in a single \emph{inseparable} step, which is inapplicable to our \emph{split} AT design. Hence, we devise a new training algorithm that combines the fast gradient approximation~\cite{ADV_01} and the variational Bayes~\cite{VAE} to learn $p(\bm{\delta}_i|\mathbf{x}_i)$ and $\Theta$ more flexibly.

\section{Methodology}
\label{sec:method}

\subsection{Motivation}
\label{subsec:motivation}

We first explain our motivation for designing SVAT. From the perspective of investors, the main source of stock recommendation risks comes from that the model may incorrectly assign higher scores to risky stocks with losses ($r_{i,t} < 0$) than profitable stocks ($r_{i,t} > 0$), after which risky stocks are recommended to investors, leading to high volatility of investment returns. One way to mitigate this problem is to train the recommendation model to be more fond of profitable stocks and more alert/sensitive to risky stocks. Adversarial training (AT) paves the way to obtain this ``split" behavior since we can indirectly manipulate the model's sensitivity to different \emph{individual} examples through their adversarial perturbations~\cite{ADV_04,ADV_02}. While conventional AT methods aim to encourage the model to be robust to imperceptible perturbations~\cite{ADV_01,ADV_02}, researchers have found that increasing the sensitivity to perturbations could also help the model better capture the diversity of data samples, calling the Inverse Adversarial Training (IAT)~\cite{IAT}. Accordingly, we posticulate that the combination of AT and IAT, namely the split AT shown in Figure~\ref{fig:split_AT}, can help the model better discriminate between profitable and risky stocks by learning from their perturbations in a different way and thus further reduce the probability of recommending risky stocks. This is the main reason why we design two different perturbations for profitable stocks and risky stocks, respectively. In addition, Variational Autoencoder (VAE)~\cite{VAE} is excellent in learning data distribution and can be used to model various stock factors~\cite{FactorVAE}. All considerations above finally converge to our SVAT method.

Although SVAT is designed to reduce stock recommendation risks, we believe that similar idea could also be applied to reduce ranking uncertainties of other learning to rank problems, such as recommender systems where positive items are more preferred than negative items.

\subsection{Overview}
\label{subsec:model_overview}

For better illustration of the risk modeling w.r.t each stock example, we literally decompose the ranking model $f$ in Equation~(\ref{equ:stock_rank}) into $N$ ranking submodules $f_1, f_2, ..., f_N$ sharing the same parameters $\Theta$, with each $f_i$ predicting the ranking score of stock $s_i$:
\begin{align}
\begin{split}
\hat{y}_{i,t} &= f_i(\tilde{\mathbf{x}}_i; \Theta),  \\
\tilde{\mathbf{x}}_i &= \Psi(X_i), \label{equ:submodules}
\end{split}
\end{align}
where $\tilde{\mathbf{x}}_i \in \mathbb{R}^D$ is the feature vector transformed from the historical sequential features $X_i = [\mathbf{x}_{i,t-T}; ...; \mathbf{x}_{i,t-1}] \in \mathbb{R}^{T \times d}$ of stock $s_i$, and $\Psi$ is the transformation function which could be simple row concatenation or temporal embedding with RNN architectures~\cite{adv_LSTM,STHAN-SR}. Similar to ADT, we model the adversarial perturbations around each stock example $\tilde{\mathbf{x}}_i$ by a conditional distribution $p(\bm{\delta}_i|\tilde{\mathbf{x}}_i)$, whose support is contained in $\Delta = \{\bm{\delta} \in \mathbb{R}^D: ||\bm{\delta}||_2 \le \epsilon\}$\footnote{We empirically found that the $l_2$-norm constraint is better for the stock recommendation problem.}. Next, we can sample a perturbation $\bm{\delta}_i$ from $p(\bm{\delta}_i|\tilde{\mathbf{x}}_i)$ to construct an adversarial example $\tilde{\mathbf{x}}_i + \bm{\delta}_i$, and obtain the perturbed output by
\begin{equation}
\hat{y}_{i,t}^{\text{adv}} = f_i(\tilde{\mathbf{x}}_i + \bm{\delta}_i; \Theta). \label{equ:adv_output}
\end{equation}

During the training phase, the adversarial loss for stock $s_i$ can be computed as
\begin{equation}
\mathcal{L}^{\text{adv}}_i = (\hat{y}_{i,t}^{\text{adv}} - y_{i,t})^2 + \alpha \sum_{j=1}^N \max \{0, -(\hat{y}_{i,t}^{\text{adv}} - \hat{y}_{j,t}^{\text{adv}})(y_{i,t} - y_{j,t})\}, \label{equ:adv_loss_i}
\end{equation}
and the total adversarial loss is the sum of each $\mathcal{L}^{\text{adv}}_i$ weighted by the corresponding stock's return ratio $r_{i,t}$:
\begin{equation}
\mathcal{L}^{\text{adv}} = \sum_{i=1}^N r_{i,t} \mathcal{L}^{\text{adv}}_i. \label{equ:adv_loss}
\end{equation}
When we train the model by minimizing $\mathcal{L}^{\text{adv}}$, the adversarial loss of stock examples with $r_{i,t} > 0$ is minimized while the adversarial loss of stock examples with $r_{i,t} < 0$ is maximized. In this way, the stock recommendation model is encouraged to be more robust to adversarial perturbations of \emph{profitable} stock examples while more sensitive to adversarial perturbations of \emph{risky} stock examples. This split adversarial training approach better enhances the risk awareness of the model by treating adversarial examples of profitable and risky stocks in an opposite way, which is consistent with the split behavior we described in the previous section.

Besides, when deploying the model to the testing environment, we can quantify the risk of each stock example by sampling multiple adversarial perturbations from $p(\bm{\delta}_i|\tilde{\mathbf{x}}_i)$. Specifically, for each testing example $\tilde{\mathbf{x}}_i$, we generate $M$ Monte Carlo perturbation samples $\bm{\delta}_i^1, \bm{\delta}_i^2, ..., \bm{\delta}_i^M$ from $p(\bm{\delta}_i|\tilde{\mathbf{x}}_i)$, and obtain different perturbed ranking scores $\hat{y}_{i,t}^{1}, \hat{y}_{i,t}^{2}, ..., \hat{y}_{i,t}^{M}$ from Equation~(\ref{equ:adv_output}). Comparing these $M$ scores with other stock examples produces $M$ rankings $a_{i,t}^1, a_{i,t}^2, ..., a_{i,t}^M$, which is used to compute the ranking entropy for $\tilde{\mathbf{x}}_i$:
\begin{equation}
\mathcal{H}(\tilde{\mathbf{x}}_i) = -\sum_{l=1}^N p(a_{i,t} = l) \cdot \log p(a_{i,t} = l), \label{equ:rank_entropy}
\end{equation}
where $p(a_{i,t} = l) = \frac{1}{M} \sum_{m=1}^M \mathbb{I}(a_{i,t}^m = l)$ denotes the frequency of $\tilde{\mathbf{x}}_i$ being ranked the $l$-th stock and we have $\mathcal{H}(\tilde{\mathbf{x}}_i) \in [0, +\infty)$. Investors could roughly evaluate the risk of the current stock example according to the ranking entropy, where higher entropy generally indicates higher risk.

\begin{figure}[t]
\centering

\subfloat[]{ 
  \begin{minipage}[t]{0.5\columnwidth}
  \centering
  \includegraphics[width=0.88\columnwidth]{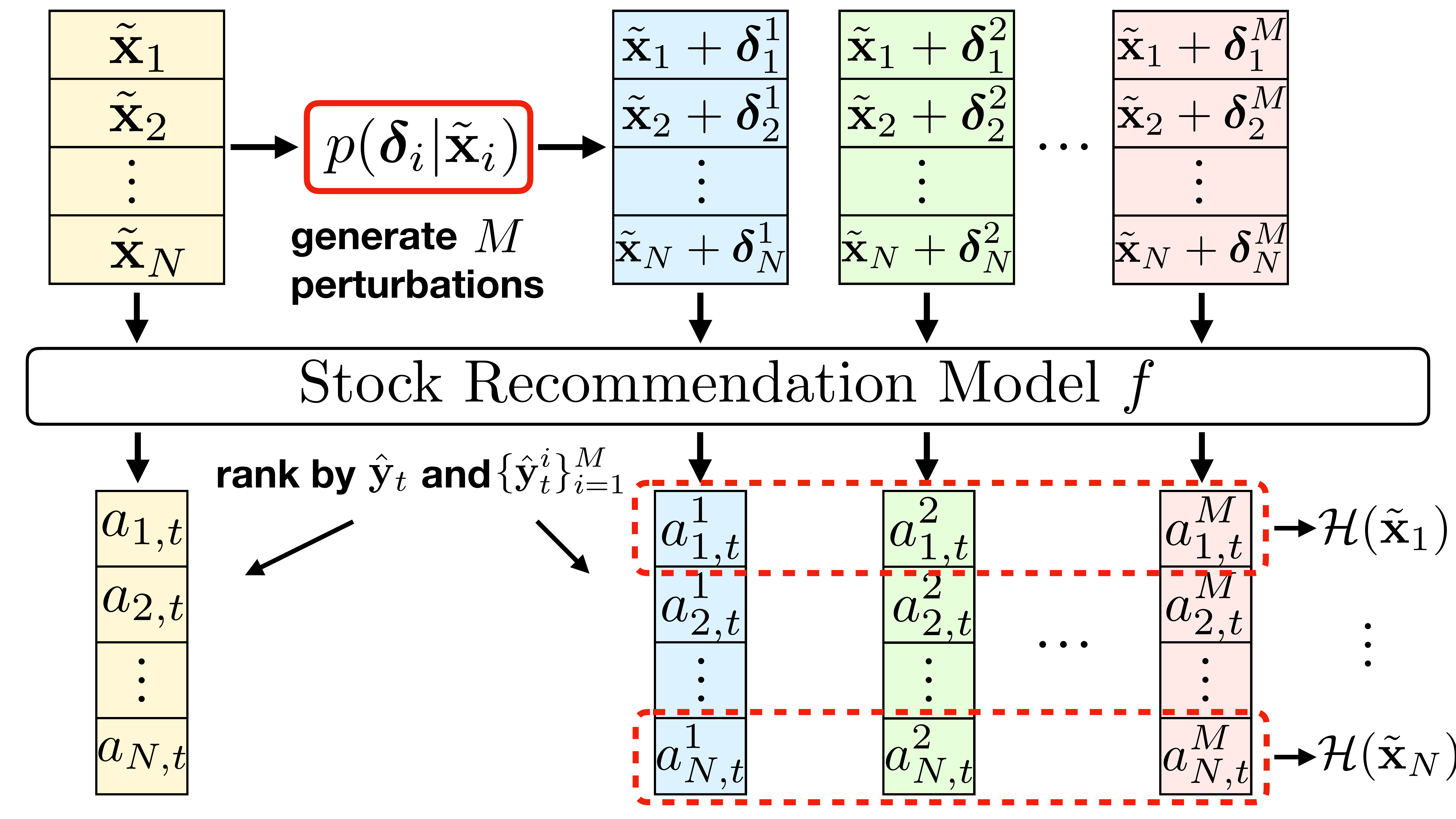}
  \label{fig:SVAT}
  \end{minipage}
}
\subfloat[]{
  \begin{minipage}[t]{0.5\columnwidth}
  \centering
  \includegraphics[width=0.88\columnwidth]{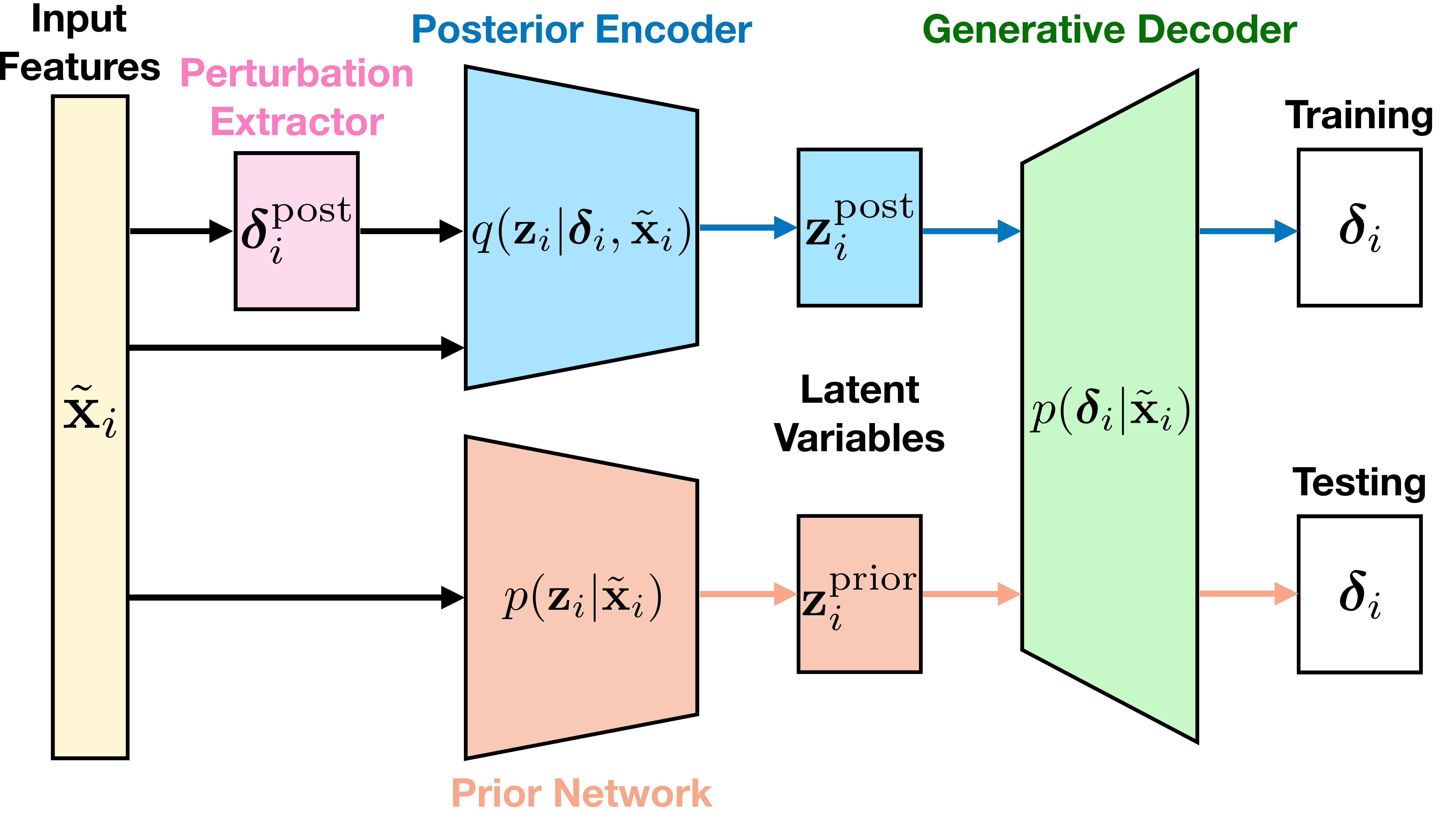}
  \label{fig:VPG}
  \end{minipage}
}
\caption{(a) Workflow of the proposed SVAT framework. (b) Architecture of the variational perturbation generator. The perturbation $\bm{\delta}_i$ generated from $\mathbf{z}_i^{\text{post}}$ is engaged in adversarial training while $\bm{\delta}_i$ generated from $\mathbf{z}_i^{\text{prior}}$ is utilized to compute the ranking entropy in the testing environment.}
\Description{SVAT framework and VPG.}
\label{fig:SVAT_VPG}
\end{figure}

Figure~\ref{fig:SVAT} presents the workflow of our SVAT method. The core of SVAT lies in the learning of the perturbation distribution $p(\bm{\delta}_i|\tilde{\mathbf{x}}_i)$, which is detailed in the next section.

\subsection{Variational Perturbation Generator}

The key role of the perturbation distribution $p(\bm{\delta}_i|\tilde{\mathbf{x}}_i)$ is to characterize potential risk factors of stocks and generate representative perturbation samples to help reduce investment risks. Since stock returns are typically affected by a variety of risk factors (e.g., macroeconomics, financial news), we decide to model these factors by some learnable latent variables that implicitly drive $p(\bm{\delta}_i|\tilde{\mathbf{x}}_i)$:
\begin{equation}
p(\bm{\delta}_i|\tilde{\mathbf{x}}_i) = \int p(\bm{\delta}_i, \mathbf{z}_i|\tilde{\mathbf{x}}_i) ~d\mathbf{z}_i = \int p(\bm{\delta}_i|\mathbf{z}_i, \tilde{\mathbf{x}}_i) p(\mathbf{z}_i|\tilde{\mathbf{x}}_i) ~d\mathbf{z}_i, \label{equ:post_infer}
\end{equation}
where $\mathbf{z}_i \in \mathbb{R}^H$ is a learnable random vector containing $H$ risk-relevant latent variables. Although the integral of the marginal likelihood in Equation~(\ref{equ:post_infer}) is intractable, it can be approximated by VAE~\cite{VAE}, a reliable framework for \emph{neural approximation}. In this spirit, we devise a Variational Perturbation Generator (VPG) to approximate $p(\bm{\delta}_i|\tilde{\mathbf{x}}_i)$ with latent factors. As shown in Figure~\ref{fig:VPG}, VPG contains three components: perturbation extractor, posterior encoder and generative decoder.

\subsubsection{Perturbation Extractor}

We first employ the fast gradient approximation method~\cite{ADV_01,adv_LSTM} as the perturbation extractor to extract a primitive perturbation sample from the gradient of the input features $\tilde{\mathbf{x}}_i$:
\begin{equation}
\bm{\delta}_i^{\text{post}} = \epsilon \cdot \frac{\nabla_{\tilde{\mathbf{x}}_i} \mathcal{L}}{||\nabla_{\tilde{\mathbf{x}}_i} \mathcal{L}||_2}, \label{equ:post_delta}
\end{equation}
where $\mathcal{L}$ is the prediction loss in Equation~(\ref{equ:origin_loss}) and $\epsilon > 0$ is the hyperparameter. $\bm{\delta}_i^{\text{post}}$ provides the posterior information about real data to guide the approximation learning of VPG.

\subsubsection{Posterior Encoder}

The posterior encoder incorporates the primitive perturbation $\bm{\delta}_i^{\text{post}}$ and the input features $\tilde{\mathbf{x}}_i$ to learn a posterior distribution $q^{\text{post}}(\mathbf{z}_i|\bm{\delta}_i, \tilde{\mathbf{x}}_i)$, from which posterior risk-relevant latent variables $\mathbf{z}^{\text{post}}_i$ can be obtained:
\begin{align}
\begin{split}
\mathbf{h}_i^{\text{post}} &= F^{\text{post}}([\bm{\delta}_i^{\text{post}}, \tilde{\mathbf{x}}_i]; \Xi^{\text{post}}), \\
\bm{\mu}^{\text{post}}_i &= W^{\text{post}}_{\mu} \mathbf{h}_i^{\text{post}} + \mathbf{b}^{\text{post}}_{\mu}, \\ 
\bm{\sigma}^{\text{post}}_i &= s^{+}(W^{\text{post}}_{\sigma} \mathbf{h}_i^{\text{post}} + \mathbf{b}^{\text{post}}_{\sigma}), \\
\mathbf{z}^{\text{post}}_i &\sim q^{\text{post}}(\mathbf{z}_i|\bm{\delta}_i, \tilde{\mathbf{x}}_i) \triangleq \mathcal{N}(\bm{\mu}^{\text{post}}_i, \text{diag}((\bm{\sigma}^{\text{post}}_i)^2)),
\label{equ:post_distrib}
\end{split}
\end{align}
where $F^{\text{post}}$ can be any non-linear neural network such as the multi-layer perceptron (MLP), $\Xi^{\text{post}}, W^{\text{post}}_{\mu}, W^{\text{post}}_{\sigma}, \mathbf{b}^{\text{post}}_{\mu}, \mathbf{b}^{\text{post}}_{\sigma}$ are learnable parameters, $s^{+}(x) = \log(1+e^x)$ denotes the $\rm{softplus}$ activation and $\mathbf{z}^{\text{post}}_i$ is a latent vector sampled from the posterior Gaussian distribution $q^{\text{post}}(\mathbf{z}_i|\bm{\delta}_i, \tilde{\mathbf{x}}_i)$.

\subsubsection{Generative Decoder}

Finally, combining the information of the latent variables $\mathbf{z}_i$ and the input features $\tilde{\mathbf{x}}_i$, we train a generative decoder network $F^{\text{gen}}$ to approximate the desired perturbation distribution $p(\bm{\delta}_i|\tilde{\mathbf{x}}_i)$ and generate the risk-indicated perturbation $\bm{\delta}_i$:
\begin{equation}
\mathbf{g}_i = F^{\text{gen}}([\mathbf{z}_i, \tilde{\mathbf{x}}_i]; \Xi^{\text{gen}}), ~~ \bm{\delta}_i = \epsilon \cdot \frac{\mathbf{g}_i}{||\mathbf{g}_i||_2}, \label{equ:delta_gen}
\end{equation}
after which $\bm{\delta}_i$ is engaged in Equation~(\ref{equ:adv_output}) to produce an adversarial example. During the training phase we simply input $\mathbf{z}_i = \mathbf{z}^{\text{post}}_i$ to $F^{\text{gen}}$, which is however unrealizable in the testing environment since we cannot obtain the loss and gradients to extract $\bm{\delta}_i^{\text{post}}$ of testing examples without knowing their ground-truth labels. Accordingly, we further design another network $F^{\text{prior}}$ to learn a prior distribution $p^{\text{prior}}(\mathbf{z}_i|\tilde{\mathbf{x}}_i)$:
\begin{align}
\begin{split}
\mathbf{h}_i^{\text{prior}} &= F^{\text{prior}}(\tilde{\mathbf{x}}_i; \Xi^{\text{prior}}), \\
\bm{\mu}^{\text{prior}}_i &= W^{\text{prior}}_{\mu} \mathbf{h}_i^{\text{prior}} + \mathbf{b}^{\text{prior}}_{\mu}, \\
\bm{\sigma}^{\text{prior}}_i &= s^{+}(W^{\text{prior}}_{\sigma} \mathbf{h}_i^{\text{prior}} + \mathbf{b}^{\text{prior}}_{\sigma}), \\
\mathbf{z}^{\text{prior}}_i &\sim p^{\text{prior}}(\mathbf{z}_i|\tilde{\mathbf{x}}_i) \triangleq \mathcal{N}(\bm{\mu}^{\text{prior}}_i, \text{diag}((\bm{\sigma}^{\text{prior}}_i)^2)),
\label{equ:prior_distrib}
\end{split}
\end{align}
and enforce the prior distribution to approximate to the posterior distribution by minimizing
\begin{equation}
\mathcal{L}^{\text{KL}}_i = D_{\text{KL}}[q^{\text{post}}(\mathbf{z}_i|\bm{\delta}_i, \tilde{\mathbf{x}}_i) || p^{\text{prior}}(\mathbf{z}_i|\tilde{\mathbf{x}}_i)], \label{equ:kl_loss}
\end{equation}
where $D_{\text{KL}}$ is the Kullback-Leibler divergence between two distributions. In this way, we can sample multiple $\mathbf{z}^{\text{prior}}_i$s from $p^{\text{prior}}(\mathbf{z}_i|\tilde{\mathbf{x}}_i)$ and generate representative perturbations for testing examples \emph{without computing their gradients}, which facilitates the risk quantification in Equation~(\ref{equ:rank_entropy}) and improves the risk interpretability of the stock recommendation model.

\subsubsection{Explanation for VPG}

As discussed in Section~\ref{subsec:motivation}, we aim to enhance the risk awareness of the stock recommendation model by encouraging the model to learn perturbations of profitable and risky stocks in a different way. Accordingly, the main goal of VPG is to provide an effective mechanism to generate perturbations for different stocks. As shown in Figure~\ref{fig:VPG}, VPG follows an encoder-decoder architecture and generates perturbations from a latent space. Since there are a variety of complex risk factors affecting stock prices, it is feasible to encode these risk factors into a latent space. Based on the framework of Variational Autoencoder (VAE)~\cite{VAE}, we assume that the perturbation $\bm{\delta}_i$ of each stock can be generated from a latent random variable $\mathbf{z}_i$ in the risk-factor latent space. We first use an encoder to learn the posterior distribution of $\mathbf{z}_i$ given $\bm{\delta}_i$ and then employ a decoder to learn a perturbation distribution close to the posterior distribution during the training phase. Finally in the testing phase, we could sample representative perturbations of all stocks from the perturbation distribution efficiently without the overhead of computing the gradient of each stock example.

In summary, the VPG can characterize various risk factors in a latent space and generate representative perturbation samples of different stocks, which is critical to enhance the risk awareness of the stock recommendation model.

\subsubsection{Theoretical Justification}\label{subsubsec:VPG_VAE}

We present the theoretical justification of VPG based on the theoretical framework of Variational Autoencoder (VAE)~\cite{VAE}. Given $N$ datapoints $\tilde{\mathbf{x}}_1, \tilde{\mathbf{x}}_2, ..., \tilde{\mathbf{x}}_N$ from the training dataset, the goal of VPG is to maximize the sum of the marginal likelihoods $\sum_{i=1}^N \log p(\bm{\delta}_i|\tilde{\mathbf{x}}_i)$, where $p(\bm{\delta}_i|\tilde{\mathbf{x}}_i)$ is driven by some risk-relevant latent variables $\mathbf{z}_i$:
\begin{align}
\begin{split}
p(\bm{\delta}_i|\tilde{\mathbf{x}}_i) &= \int p(\bm{\delta}_i, \mathbf{z}_i|\tilde{\mathbf{x}}_i)d\mathbf{z}_i \\ 
&= \int p(\bm{\delta}_i|\mathbf{z}_i, \tilde{\mathbf{x}}_i) p(\mathbf{z}_i|\tilde{\mathbf{x}}_i)d\mathbf{z}_i \\
&= \int p(\bm{\delta}_i|\mathbf{z}_i, \tilde{\mathbf{x}}_i) p(\mathbf{z}_i|\tilde{\mathbf{x}}_i)d\mathbf{z}_i \\
&= \mathbb{E}_{\mathbf{z}_i \sim p(\mathbf{z}_i|\tilde{\mathbf{x}}_i)}[p(\bm{\delta}_i|\mathbf{z}_i, \tilde{\mathbf{x}}_i)]. \label{equ:VPG_expect}
\end{split}
\end{align}
Without loss of generality, we assume that $p(\mathbf{z}_i|\tilde{\mathbf{x}}_i)$ is a Gaussian distribution conditioned on $\tilde{\mathbf{x}}_i$ and $p(\bm{\delta}_i|\mathbf{z}_i, \tilde{\mathbf{x}}_i)$ is a Gaussian distribution conditioned on both $\tilde{\mathbf{x}}_i$ and $\mathbf{z}_i$:
\begin{align}
\begin{split}
p(\mathbf{z}_i|\tilde{\mathbf{x}}_i) &\triangleq \mathcal{N}(\mu_{\mathbf{z}}(\tilde{\mathbf{x}}_i), \Sigma_{\mathbf{z}}(\tilde{\mathbf{x}}_i)), \\
p(\bm{\delta}_i|\mathbf{z}_i, \tilde{\mathbf{x}}_i) &\triangleq \mathcal{N}(\mu_{\bm{\delta}}(\mathbf{z}_i, \tilde{\mathbf{x}}_i), \Sigma_{\bm{\delta}}(\mathbf{z}_i, \tilde{\mathbf{x}}_i)),
\end{split}
\end{align}
where $\mu_{\mathbf{z}}(\cdot), \mu_{\bm{\delta}}(\cdot, \cdot)$ are learnable functions that output the mean of the Gaussian and $\Sigma_{\mathbf{z}}(\cdot), \Sigma_{\bm{\delta}}(\cdot, \cdot)$ are learnable functions that output the covariance of the Gaussian, all of which can be approximated by neural networks. In this case, we can sample a large number of $\mathbf{z}_i$ from $p(\mathbf{z}_i|\tilde{\mathbf{x}}_i)$ and approximate the expectation in Equation~(\ref{equ:VPG_expect}) by average:
\begin{align}
\begin{split}
p(\bm{\delta}_i|\tilde{\mathbf{x}}_i) = \mathbb{E}_{\mathbf{z}_i \sim p(\mathbf{z}_i|\tilde{\mathbf{x}}_i)}&[p(\bm{\delta}_i|\mathbf{z}_i, \tilde{\mathbf{x}}_i)] \approx \frac{1}{K} \sum_{k=1}^K \mathcal{N}(\mu_{\bm{\delta}}(\mathbf{z}_i^k, \tilde{\mathbf{x}}_i), \Sigma_{\bm{\delta}}(\mathbf{z}_i^k, \tilde{\mathbf{x}}_i)), \\
\text{where}&~~ \mathbf{z}_i^k \sim \mathcal{N}(\mu_{\mathbf{z}}(\tilde{\mathbf{x}}_i), \Sigma_{\mathbf{z}}(\tilde{\mathbf{x}}_i)). \label{equ:sample_z}
\end{split}
\end{align}
However, this approach suffers from \emph{curse of dimensionality} since the sample number $K$ grows exponentially as the dimension of $\mathbf{z}_i$ increases. Besides, for any given observations $\bm{\delta}_i$ and $\tilde{\mathbf{x}}_i$, most $\mathbf{z}_i^k$ will contribute very little to the likelihood.

To solve the problem above, Variational Autoencoder (VAE)~\cite{VAE} proposes to sample the latent variables $\mathbf{z}_i$ from the posterior distribution $p(\mathbf{z}_i|\bm{\delta}_i, \tilde{\mathbf{x}}_i)$ and only pick \emph{a small number of $\mathbf{z}_i$ values} that contribute a significant amount to the likelihood. Specifically, VAE approximates the ground-truth posterior distribution $p(\mathbf{z}_i|\bm{\delta}_i, \tilde{\mathbf{x}}_i)$ by a learnable model $q_{\bm{\phi}}(\mathbf{z}_i|\bm{\delta}_i, \tilde{\mathbf{x}}_i)$ with parameters $\bm{\phi}$. Considering the Kullback-Leibler divergence between $q_{\bm{\phi}}(\mathbf{z}_i|\bm{\delta}_i, \tilde{\mathbf{x}}_i)$ and $p(\mathbf{z}_i|\bm{\delta}_i, \tilde{\mathbf{x}}_i)$, we obtain:
\begin{align}
\begin{split}
&D_{\text{KL}}(q_{\bm{\phi}}(\mathbf{z}_i|\bm{\delta}_i, \tilde{\mathbf{x}}_i) || p(\mathbf{z}_i|\bm{\delta}_i, \tilde{\mathbf{x}}_i)) \\
\triangleq &\int q_{\bm{\phi}}(\mathbf{z}_i|\bm{\delta}_i, \tilde{\mathbf{x}}_i) \log [\frac{q_{\bm{\phi}}(\mathbf{z}_i|\bm{\delta}_i, \tilde{\mathbf{x}}_i)}{p(\mathbf{z}_i|\bm{\delta}_i, \tilde{\mathbf{x}}_i)}] d \mathbf{z}_i \\
= &\mathbb{E}_{\mathbf{z}_i \sim q_{\bm{\phi}}}[\log q_{\bm{\phi}}(\mathbf{z}_i|\bm{\delta}_i, \tilde{\mathbf{x}}_i) - \log p(\mathbf{z}_i|\bm{\delta}_i, \tilde{\mathbf{x}}_i)] \\
= &\mathbb{E}_{\mathbf{z}_i \sim q_{\bm{\phi}}}[\log q_{\bm{\phi}}(\mathbf{z}_i|\bm{\delta}_i, \tilde{\mathbf{x}}_i) - \log \frac{p(\bm{\delta}_i|\mathbf{z}_i, \tilde{\mathbf{x}}_i) p(\mathbf{z}_i|\tilde{\mathbf{x}}_i)}{p(\bm{\delta}_i|\tilde{\mathbf{x}}_i)}] ~~(\text{Bayes rule}) \\
= &\mathbb{E}_{\mathbf{z}_i \sim q_{\bm{\phi}}}[\log q_{\bm{\phi}}(\mathbf{z}_i|\bm{\delta}_i, \tilde{\mathbf{x}}_i) - \log p(\bm{\delta}_i|\mathbf{z}_i, \tilde{\mathbf{x}}_i) - \log p(\mathbf{z}_i|\tilde{\mathbf{x}}_i)] + \log p(\bm{\delta}_i|\tilde{\mathbf{x}}_i)\\
= &\log p(\bm{\delta}_i|\tilde{\mathbf{x}}_i) -\mathbb{E}_{\mathbf{z}_i \sim q_{\bm{\phi}}}[\log p(\bm{\delta}_i|\mathbf{z}_i, \tilde{\mathbf{x}}_i)] +\mathbb{E}_{\mathbf{z}_i \sim q_{\bm{\phi}}}[\log q_{\bm{\phi}}(\mathbf{z}_i|\bm{\delta}_i, \tilde{\mathbf{x}}_i) - \log p(\mathbf{z}_i|\tilde{\mathbf{x}}_i)] \\
= &\log p(\bm{\delta}_i|\tilde{\mathbf{x}}_i) -\mathbb{E}_{\mathbf{z}_i \sim q_{\bm{\phi}}}[\log p(\bm{\delta}_i|\mathbf{z}_i, \tilde{\mathbf{x}}_i)] + D_{\text{KL}}[q_{\bm{\phi}}(\mathbf{z}_i|\bm{\delta}_i, \tilde{\mathbf{x}}_i) || p(\mathbf{z}_i|\tilde{\mathbf{x}}_i)] \label{equ:KL_derive}
\end{split}
\end{align}
Rearranging Equation~(\ref{equ:KL_derive}), we finally obtain the \emph{evidence lower bound (ELBO)} proposed in~\cite{VAE}:
\begin{align}
\begin{split}
&\log p(\bm{\delta}_i|\tilde{\mathbf{x}}_i) - D_{\text{KL}}[q_{\phi}(\mathbf{z}_i|\bm{\delta}_i, \tilde{\mathbf{x}}_i) || p(\mathbf{z}_i|\bm{\delta}_i, \tilde{\mathbf{x}}_i)] \\
= &\mathbb{E}_{\mathbf{z}_i \sim q_{\phi}}[\log p(\bm{\delta}_i|\mathbf{z}_i, \tilde{\mathbf{x}}_i)] - D_{\text{KL}}[q_{\phi}(\mathbf{z}_i|\bm{\delta}_i, \tilde{\mathbf{x}}_i) || p(\mathbf{z}_i|\tilde{\mathbf{x}}_i)]. \label{equ:VAE}
\end{split}
\end{align}
Therefore, we can maxmize the marginal likelihood $\log p(\bm{\delta}_i|\tilde{\mathbf{x}}_i)$ and minimize the Kullback-Leibler divergence $D_{\text{KL}}[q_{\phi}(\mathbf{z}_i|\bm{\delta}_i, \tilde{\mathbf{x}}_i) || p(\mathbf{z}_i|\bm{\delta}_i, \tilde{\mathbf{x}}_i)]$ (i.e., the LHS of Equation~(\ref{equ:VAE})) by equivalently maximizing $\mathbb{E}_{\mathbf{z}_i \sim q_{\phi}}[\log p(\bm{\delta}_i|\mathbf{z}_i, \tilde{\mathbf{x}}_i)]$ and minimizing $D_{\text{KL}}[q_{\phi}(\mathbf{z}_i|\bm{\delta}_i, \tilde{\mathbf{x}}_i) || p(\mathbf{z}_i|\tilde{\mathbf{x}}_i)]$ (i.e., the RHS of Equation~(\ref{equ:VAE})), which is exactly what VPG does. As shown in Figure~\ref{fig:VPG}, VPG utilizes the Posterior Encoder and the Prior Network to model the approximated posterior distribution $q_{\phi}(\mathbf{z}_i|\bm{\delta}_i, \tilde{\mathbf{x}}_i)$ and the prior distribution $p(\mathbf{z}_i|\tilde{\mathbf{x}}_i)$, respectively. And we minimize their Kullback-Leibler divergence by minimizing the $\mathcal{L}_i^{\text{KL}}$ in Equation~(\ref{equ:kl_loss}). On the other hand, since all perturbations $\bm{\delta}_i$ generated from $p(\bm{\delta}_i|\mathbf{z}_i, \tilde{\mathbf{x}}_i)$ are expected to be representative risk indicators that can minimize the $\mathcal{L}^{\text{adv}}$ in Equation~(\ref{equ:adv_loss}), we can maximize the likelihood $\mathbb{E}_{\mathbf{z}_i \sim q_{\phi}}[\log p(\bm{\delta}_i|\mathbf{z}_i, \tilde{\mathbf{x}}_i)]$ by equivalently minimizing the $\mathcal{L}^{\text{adv}}$. Finally, we conclude that the proposed SVAT loss $\mathcal{L}^{\text{adv}} + \mathcal{L}^{\text{KL}}$ is consistent with the theoretical framework of VAE.

Note that during the training phase, we only sample one $\mathbf{z}_i$ from the approximated posterior distribution $q_{\phi}(\mathbf{z}_i|\bm{\delta}_i, \tilde{\mathbf{x}}_i)$ in each epoch and approximate the expectation $\mathbb{E}_{\mathbf{z}_i \sim q_{\phi}}[\log p(\bm{\delta}_i|\mathbf{z}_i, \tilde{\mathbf{x}}_i)]$ by training the VPG for multiple epochs, thus avoiding a large number of sampling as in Equation~(\ref{equ:sample_z}) and the curse of dimensionality.

\subsection{Model Training}

We summarize the training process of SVAT as Algorithm~\ref{alg:SVAT}. The stock recommendation model $f$ and all components of the VPG are trained end-to-end by minimizing the combined loss function of Equation~(\ref{equ:origin_loss}),~(\ref{equ:adv_loss}) and~(\ref{equ:kl_loss}):
\begin{equation}
\mathcal{L}^{\text{com}} = \mathcal{L} + \lambda(\mathcal{L}^{\text{adv}} + \mathcal{L}^{\text{KL}}), \nonumber
\end{equation}
where $\mathcal{L}^{\text{KL}} = \sum_{i=1}^N \mathcal{L}^{\text{KL}}_i$ and $\lambda$ is a hyperparameter to control the contribution of the SVAT loss. We utilize the Adam~\cite{adam} algorithm for optimization.

\begin{algorithm}[t]
\renewcommand{\algorithmicrequire}{\textbf{Input:}}
\renewcommand{\algorithmicensure}{\textbf{Output:}}
\caption{Split Variational Adversarial Training}
\label{alg:SVAT}
\begin{algorithmic}[1]
\REQUIRE Stock set $\mathcal{S} = \{s_i\}_{i=1}^N$, training data $\mathcal{D}$, training epochs $E$, learning rate $\eta$, and hyperparamters $\epsilon$, $\alpha$, $\lambda$. \\
\STATE Initialize model parameters $\Theta$ and $\Xi = \{\Xi^{\text{post}}, \Xi^{\text{gen}}, \Xi^{\text{prior}}, W^{\text{post}}_{\mu}, W^{\text{post}}_{\sigma}, W^{\text{prior}}_{\mu}, W^{\text{prior}}_{\sigma}, \mathbf{b}^{\text{post}}_{\mu}, \mathbf{b}^{\text{post}}_{\sigma},$ \\ $\mathbf{b}^{\text{prior}}_{\mu}, \mathbf{b}^{\text{prior}}_{\sigma}\}$;
\FOR{epoch $= 1$ \textbf{to} $E$}
\FOR{each batch input $(\mathbf{X}, \mathbf{y}_t) \in \mathcal{D}$}
\STATE Predict ranking scores $\mathbf{\hat{y}}_t$ by Equation~(\ref{equ:stock_rank});
\STATE Obtain $\mathcal{L}$ by Equation~(\ref{equ:origin_loss});
\STATE Obtain $\{\bm{\delta}^{\phi}_i\}_{i=1}^N$ by Equation~(\ref{equ:submodules}) and~(\ref{equ:post_delta});
\STATE Obtain $\{\mathbf{z}_i\}_{i=1}^N$ by Equation~(\ref{equ:post_distrib}) and~(\ref{equ:prior_distrib});
\STATE Generate perturbations $\{\bm{\delta}_i\}_{i=1}^N$ by Equation~(\ref{equ:delta_gen});
\STATE Predict adversarial scores $\mathbf{\hat{y}}_t^{\text{adv}}$ by Equation~(\ref{equ:adv_output});
\STATE Update $\Theta \gets \Theta - \eta \cdot \nabla_{\Theta} \mathcal{L}^{\text{com}}$; 
\STATE Update $\Xi \gets \Xi - \eta \cdot \nabla_{\Xi} \mathcal{L}^{\text{com}}$; 
\ENDFOR
\ENDFOR
\ENSURE A risk-aware stock model $f$ and a variational perturbation generator $G_{\bm{\delta}}$.
\end{algorithmic}
\end{algorithm}

\section{Experiments}
\label{sec:experiments}

The core of this section is to evaluate whether our method can effectively reduce stock investment risks for investors. Accordingly, we conduct extensive experiments with the aim of answering the following research questions:
\begin{itemize}
  \item \textbf{RQ1:} How is the utility of our proposed SVAT method in a general economic environment? Can SVAT outperform state-of-the-art stock recommendation models in terms of risk-adjusted profits under normal circumstances?
  \item \textbf{RQ2:} How is the utility of our SVAT method in an extreme economic environment such as the financial crisis? Can SVAT protect investors from risks better than other state-of-the-art stock recommendation models under extreme circumstances?
  \item \textbf{RQ3:} Does SVAT capture different signals or recommend stocks different from other baseline methods? To what extent are all methods correlated with each other?
  \item \textbf{RQ4:} How is the effectiveness of the split adversarial training design and the variational perturbation generator component of our proposed SVAT method?
  \item \textbf{RQ5:} How does our proposed SVAT method perform under different backtesting strategies, different adversarial hyperparameter settings and different sampling methods?
  \item \textbf{RQ6:} How does the adversarial perturbation of SVAT help reduce the risk of stock recommendation? What insights can investors learn from SVAT?
\end{itemize}
We next conduct different experiments to answer the RQs above, comprehensively demonstrating the effectiveness, practicality, and robustness of our approach.

\subsection{Experimental Setting}

\subsubsection{Datasets}

As shown in Table~\ref{tab:data_statistics}, our experiments are based on six real-world datasets from \emph{US} and \emph{China} stock markets, including three \emph{normal datasets} in a general economic environment (RQ1) and three \emph{crisis datasets} during the financial crisis period (RQ2):
\begin{itemize}
  \item \emph{Normal datasets}:
  \begin{itemize}
    \item \textbf{NASDAQ}~\cite{RSR}: This dataset contains the price data of $1,026$ equity stocks in the NASDAQ Global and Capital market from 01/02/2013 to 12/08/2017.
    \item \textbf{NYSE}~\cite{RSR}: This dataset consists of the price data of $1,737$ equity stocks in the New York Stock Exchange market from 01/02/2013 to 12/08/2017.
    \item \textbf{CASE}: This dataset collects the price data of $4,465$ equity stocks from the China A-share Stock Exchange market from 03/01/2016 to 03/04/2022.
  \end{itemize}
  \item \emph{Crisis datasets}:
  \begin{itemize}
    \item \textbf{NASDAQ\_08}: This dataset contains the price data of $656$ equity stocks in the NASDAQ Global and Capital market from 01/02/2002 to 12/31/2008.
    \item \textbf{NYSE\_08}: This dataset consists of the price data of $1,115$ equity stocks in the New York Stock Exchange market from 01/02/2002 to 12/31/2008.
    \item \textbf{CASE\_08}: This dataset collects the price data of $1,520$ equity stocks from the China A-share Stock Exchange market from 01/04/2002 to 12/31/2008.
  \end{itemize}
\end{itemize}
All the stock datasets above are collected every one day (i.e., daily price data), within which each data point consists of 5 features (i.e., the opening price, highest price, lowest price, closing price, and trading volume of the stock for the day). Among these datasets, \textbf{NASDAQ} and \textbf{NYSE} have been widely used in most previous work~\cite{RSR,STHAN-SR,ALSP_TF} and we also use them here to ensure fair comparison with other state-of-the-art models. On the other hand, we collect the data of \textbf{CASE} and \textbf{CASE\_08} from RiceQuant\footnote{\url{https://www.ricequant.com/}}, the data of \textbf{NASDAQ\_08} and \textbf{NYSE\_08} from Yahoo Finance\footnote{\url{https://finance.yahoo.com/}}, respectively. In particular, \textbf{NASDAQ\_08}, \textbf{NYSE\_08}, and \textbf{CASE\_08} datasets contain the stock data covering the entire 2007-2008 global financial crisis period, which is crucial to test the anti-risk ability of stock prediction models.

\begin{table}[htbp]
  \centering
  \caption{Dataset statistics detailing chronological date splits of the six stock datasets.}
  \label{tab:data_statistics}
  \resizebox{0.9\linewidth}{!}{
  \begin{tabular}{c||c|c|c|c|c|c}
    \toprule
    Dataset            & NASDAQ           & NYSE            & CASE            & NASDAQ\_08      & NYSE\_08        & CASE\_08         \\
    \toprule
    Train(Tr) Period   & 01/2013-12/2015  & 01/2013-12/2015 & 03/2016-04/2019 & 01/2002-12/2006 & 01/2002-12/2006 & 01/2002-12/2006  \\
    \midrule
    Valid(Va) Period   & 01/2016-12/2016  & 01/2016-12/2016 & 04/2019-04/2020 & 01/2007-10/2007 & 01/2007-10/2007 & 01/2007-10/2007  \\
    \midrule
    Test(Te) Period    & 01/2017-12/2017  & 01/2017-12/2017 & 04/2020-03/2022 & 11/2007-12/2008 & 11/2007-12/2008 & 11/2007-12/2008  \\
    \midrule
    $\#$Days(Tr:Va:Te) & 756:252:237      & 756:252:237     & 756:252:456     & 1259:211:295    & 1259:211:295    & 1205:201:289     \\
    \midrule
    $\#$Stocks         & $1,026$          & $1,737$         & $4,465$         & $656$           & $1,115$         & $1,520$          \\
    \bottomrule
  \end{tabular}
  }
\end{table}

\subsubsection{Baselines}

Since we focus on reducing investment risks of stock recommendation, we compare our method with the stock market composite index and 7 stock recommendation baseline methods as follows:
\begin{itemize}
    \item \textbf{Buy\&Hold}: This is the simplest trading strategy where we buy the composite index of all stocks and hold. The results of this buy and hold index represent the average benchmark performance of the stock market.
    \item \textbf{ARIMA}~\cite{ARIMA}: This method is the traditional Autoregressive Integrated Moving Average model for time series prediction. We use it to directly predict the return ratio of each stock and recommend stocks with the highest predicted return ratios.
    \item \textbf{LSTM}~\cite{stock_LSTM}: This method is the vanilla LSTM model which operates on the sequential stock price data and obtains a sequential embedding for stock recommendation. We finally combine it with a fully-connected layer to predict the ranking score of each stock.
    \item \textbf{GCN}~\cite{GCN}: GCN is the typical and representative graph-based learning method. We use the vanilla GCN architecture to model the stock relation graph and combine it with a fully-connected layer to predict the ranking score of each stock.
    \item \textbf{RSR-E}~\cite{RSR}: This method develops a temporal GCN model using price movement similarity to weight the relation between different stocks and improves the performance of stock relation learning.
    \item \textbf{RSR-I}~\cite{RSR}: This method employs an implicit neural network to adatively learn the stock relation and improves the adaptability of the temporal GCN model.
    \item \textbf{ANN-SVM}~\cite{ANN-SVM}: This method incorporates a non-linear Artificial Neural Network (ANN) and a Support Vector Machine (SVM) to perform stock recommendation.
    \item \textbf{STHAN-SR}~\cite{STHAN-SR}: This method leverages a hypergraph attention network to learn the stock relation and achieve great improvements on stock recommendation.
\end{itemize}

\subsubsection{Evaluation Metrics}
\label{subsubsec:metrics}

Following the previous work~\cite{RSR,TRAN,STHAN-SR,ALSP_TF}, we also adopt a daily buy-hold-sell trading strategy and evaluate all stock recommendation methods by the following three metrics: 
\begin{enumerate}
    \item Investment Return Ratio (IRR): \begin{equation}
    \text{IRR} = \sum_t \text{IRR}^t = \sum_t \sum_{i \in \mathcal{S}^{t-1}} r_{i,t}, \nonumber
    \end{equation}
    where $\mathcal{S}^{t-1} \subseteq \mathcal{S}$ denotes the set of top-$k$ stocks selected on trading day $t-1$ and $r_{i,t}$ is the $1$-day return ratio of stock $s_i$ on trading day $t$. IRR only evaluates the model's profit without risk consideration.
    \item Sharpe Ratio (SR): \begin{equation}
    \text{SR} = \frac{\mathbb{E}[\text{IRR}^t - R_f]}{\text{STD}[\text{IRR}^t - R_f]}, \label{equ:SR}
    \end{equation}
    where $R_f$ is a risk-free return\footnote{T-Bill rates: \url{https://home.treasury.gov/}} and STD denotes the standard deviation. SR is a risk-adjusted return metric considering both the profit and the volatility of the model.
    \item Maximum Daily Drawdown (MDD): \begin{equation}
    \text{MDD} = 100\% \times |\min\{\min_t \text{IRR}^t, 0\}|. \nonumber
    \end{equation}
    MDD measures the maximum daily loss of the model in backtesting (e.g., the MDD of Model 1 in Figure~\ref{fig:return_risk} is $51.7\%$), which evaluates to what extent can the model protect investors from the risk of loss.
\end{enumerate}

\subsubsection{Training Setup}

We implement the models with \texttt{PyTorch}\footnote{https://pytorch.org/} except ARIMA of which we use the \texttt{Python statsmodels} package implementation\footnote{https://www.statsmodels.org/stable/index.html}. For fair comparison, we use grid search to select optimal hyperparameters regarding SR for each model. For all methods, we tune the length of sequential input $T$ within $\{4, 8, 16, 20\}$ and the learning rate $\eta$ within $\{1e-4, 5e-3\}$. For the LSTM and the GCN model, we tune the number of hidden units within $\{16, 32, 64, 128\}$. For RSR-E, RSR-I, ANN-SVM and STHAN-SR models, we conduct the same hyperparameter tuning as reported in their original papers~\cite{RSR,ANN-SVM,STHAN-SR}. As for our SVAT method, we employ STHAN-SR~\cite{STHAN-SR} as the backbone recommendation model and tune the adversarial constraint $\epsilon$ within $\text{range}[0.001, 0.1]$, loss weighting factors $\alpha, \lambda$ within $\text{range}[0.1, 1]$. We employ a two-layer MLP with $128$ hidden neurons and $\tanh$ activation to construct $F^{\text{post}}, F^{\text{gen}}, F^{\text{prior}}$, respectively. We set $k=5$ and thus select top-$5$ stocks each day for evluation. Finally, we train all models on a Tesla V100 GPU for $E=500$ epochs.

\subsubsection{Discussion: Predictability of Stock Daily Returns}

The core task of our SVAT method and other stock recommendation baseline methods is to predict the daily returns of each stock and perform stock ranking, which relies on the basic premise that \emph{the stock daily returns are predictable}. Indeed, there are some references~\cite{stock_predictable_01,stock_predictable_02,stock_predictable_03,stock_predictable_04} that provide reliable evidence for the predictability of stock daily returns and thus support the rationality and feasibility for current stock recommendation researches. However, most of the references above used the stock market data in 1978-2002 to produce their evidence. Since the information transparency and the speed of information spread in 1978-2002 may differ from that in the current Internet era, it is necessary to further verify the predictability of stock daily returns in the Internet era. Unfortunately, we have not yet found any references using recent data to provide relevant evidence so far. Hence, we remain cautious about the predictability of stock daily returns in the Internet era and will explore this topic further in future work.

\subsection{RQ1: Performance Comparison of Normal Economic Environment}

\begin{table}[htbp]
    \centering
    \caption{Backtesting results of the three normal datasets. $\uparrow$ means the larger the better while $\downarrow$ means the smaller the better. Among the results of all stock recommendation methods, the best result is in \textbf{boldface} and the second best result is \underline{underlined}.}
    \label{tab:result}
    \resizebox{0.75\textwidth}{!}{
    \begin{tabular}{l|ccc|ccc|ccc}
    \toprule[1pt]
    Dataset                       & \multicolumn{3}{c|}{NASDAQ}                                       & \multicolumn{3}{c|}{NYSE}                                         & \multicolumn{3}{c}{CASE}                                         \\
    \midrule[.5pt]
    Model                         & IRR$\uparrow$       & SR$\uparrow$        & MDD$\downarrow$       & IRR$\uparrow$       & SR$\uparrow$        & MDD$\downarrow$       & IRR$\uparrow$       & SR$\uparrow$        & MDD$\downarrow$      \\
    \midrule[.5pt]
    Buy\&Hold                     & $0.24$              & $2.43$              & $2.57\%$              & $0.14$              & $1.96$              & $1.58\%$              & $0.22$              & $0.71$              & $4.50\%$             \\
    \midrule[.2pt]                                                                                                                                                                                                                              
    ARIMA                         & $0.10$              & $0.55$              & $8.21\%$              & $0.10$              & $0.33$              & $7.15\%$              & $0.23$              & $0.30$              & $8.53\%$             \\
    LSTM                          & $0.22$              & $0.95$              & $7.37\%$              & $0.12$              & $0.79$              & \underline{$5.72\%$}  & $0.35$              & $0.53$              & $7.75\%$             \\
    GCN                           & $0.13$              & $0.46$              & $7.91\%$              & $0.16$              & $0.72$              & $6.20\%$              & $0.43$              & $1.03$              & $7.33\%$             \\
    RSR-E                         & $0.26$              & $1.12$              & $7.35\%$              & $0.20$              & $0.88$              & $5.86\%$              & $0.56$              & $0.96$              & \underline{$5.95\%$} \\
    RSR-I                         & $0.39$              & $1.34$              & $6.75\%$              & $0.21$              & $0.95$              & $6.34\%$              & $0.58$              & $1.04$              & $7.95\%$             \\
    ANN-SVM                       & $0.32$              & $1.28$              & \underline{$5.73\%$}  & \underline{$0.33$}  & \underline{$1.14$}  & $8.59\%$              & $0.26$              & $0.43$              & $5.97\%$             \\
    STHAN-SR                      & \underline{$0.44$}  & \underline{$1.42$}  & $6.37\%$              & \underline{$0.33$}  & $1.12$              & $6.09\%$              & \underline{$0.71$}  & \underline{$1.09$}  & $7.80\%$             \\
    SVAT (Ours)                   & $\bm{0.59}$         & $\bm{3.10}$         & $\bm{3.29\%}$         & $\bm{0.38}$         & $\bm{2.61}$         & $\bm{3.13\%}$         & $\bm{0.79}$         & $\bm{1.50}$         & $\bm{5.88\%}$        \\ 
    \bottomrule[1pt]
    \end{tabular}}
\end{table}

Table~\ref{tab:result} summarizes the experimental results of the buy and hold index and all stock recommendation methods in terms of profitability and risk on the three normal datasets (NASDAQ, NYSE, and CASE). We can obtain the following observations:
\begin{itemize}
  \item Comparing the results of Buy\&Hold with those of other stock recommendation methods, we observe that Buy\&Hold achieves the smallest MDD and competitive SR on the three normal datasets, mainly because it reduces the volatility of returns by averaging the stock returns across the whole market. However, the IRR of Buy\&Hold is more than $50\%$ less than other state-of-the-art stock recommendation models such as RSR, ANN-SVM, STHAN-SR, and our SVAT methods. Therefore, the buy and hold index cannot attain a good balance between profits and risks in the stock market and it is necessary to develop more advanced stock recommendation methods.
  \item Among all stock recommendation methods, our SVAT method outperforms the other baselines on all datasets, showing the superiority of the split variational adversarial training design for stock recommendation. Specifically, SVAT improves the SR by an average of $94.96\%$ and reduces the MDD by an average of $29.68\%$ compared to the second best results of other methods. Such an improvement answers the RQ1 that SVAT does outperform state-of-the-art stock recommendation models in terms of risk-adjusted profits under normal circumstances.
  \item As for profitability, SVAT improves the cumulative return ratio (IRR) by an average of $20.17\%$ on the three normal datasets. This shows that increasing the risk-sensitivity of the model also helps improve the investment profit, probably by encouraging the model to avoid selecting stock examples with high risks.
  \item Compared to the original backbone model STHAN-SR, SVAT greatly improves IRR, SR, MDD by an average of $20.17\%, 96.32\%, 40.52\%$, respectively, which demonstrates that our method does help the stock recommendation model to effectively reduce investment risks and control potential losses under a safer and more tolerable risk level.
  \item Without explicit risk modeling, the state-of-the-art models RSR-I, ANN-SVM and STHAN-SR even attain worse MDDs than simple methods LSTM and GCN on the NYSE and/or CASE datasets, which indicates that recent advanced stock recommendation models may not have achieved improvement in risk control. Hence, it is necessary and promising to design a method like SVAT to enhance the risk awareness of stock recommendation.
\end{itemize}

\begin{figure}[t]
\centering

\subfloat[Daily returns of all models. \texttt{STD} denotes the standard deviation of the model profit.]{ 
  \begin{minipage}[t]{\columnwidth}
  \centering
  \includegraphics[width=\columnwidth]{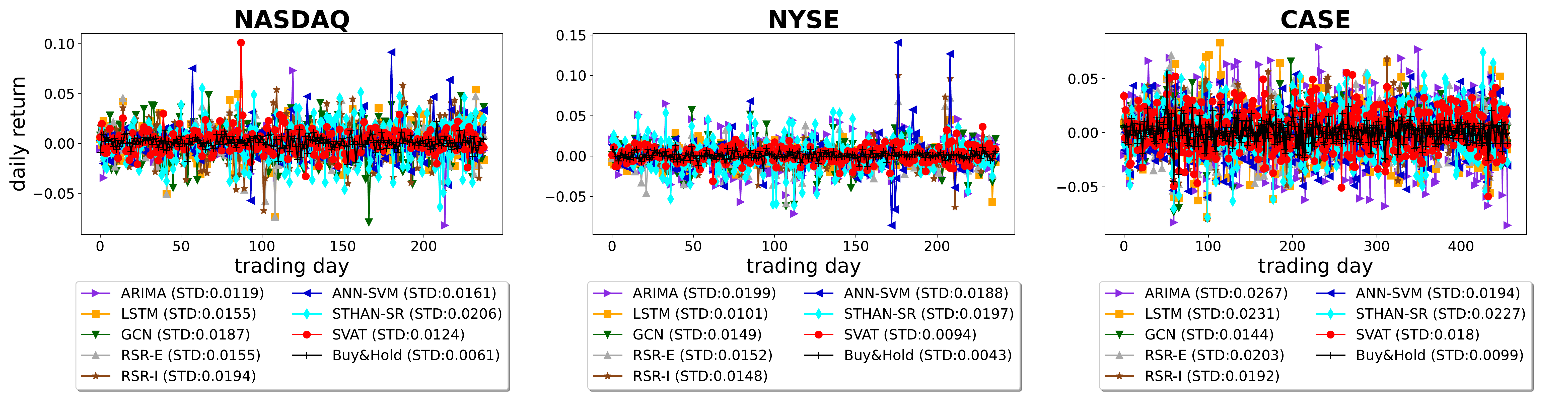}
  \label{fig:daily_returns}
  \end{minipage}
}

\subfloat[Cumulative investment return ratios (IRR) of all models.]{
  \begin{minipage}[t]{\columnwidth}
  \centering
  \includegraphics[width=\columnwidth]{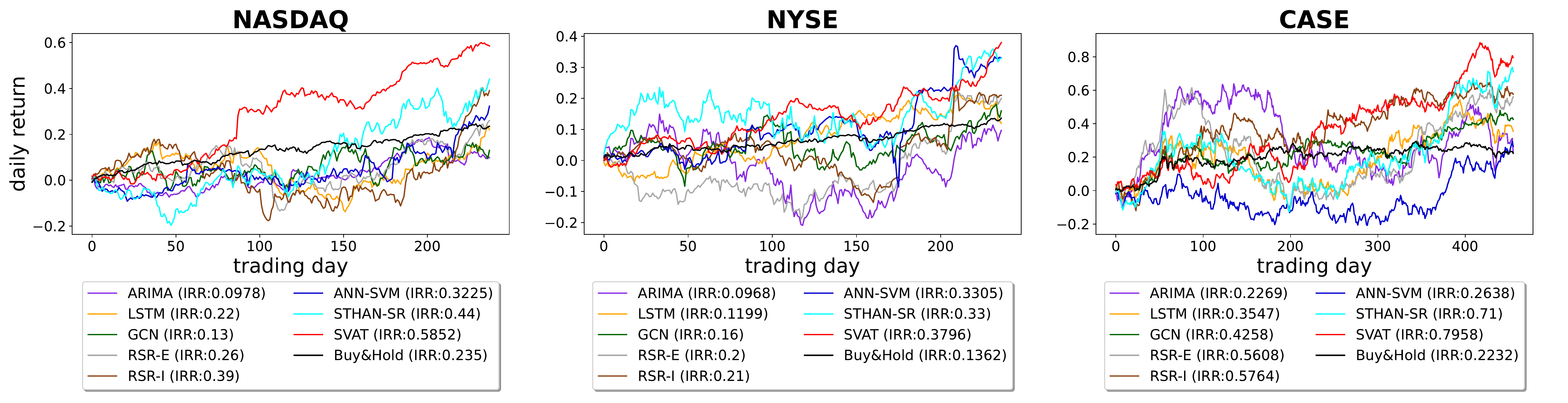}
  \label{fig:cumulative_IRR}
  \end{minipage}
}
\caption{The curves of daily returns and cumulative investment return ratios of all models backtested on the three normal datasets. Best viewed in color.}
\Description{Daily and cumulative returns of all models.}
\label{fig:returns}
\end{figure}

Figure~\ref{fig:returns} presents the curves of daily returns and the cumulative investment return ratios (IRR) of all models backtested on the three normal datasets. Clearly, the standard deviation (STD) of SVAT's daily returns is smaller and the IRR curve of SVAT is less volatile than recent advanced stock recommendation models RSR-E, RSR-I, ANN-SVM and STHAN-SR. Although the STD of SVAT's daily returns is slightly larger than simple methods ARIMA/GCN on the NASDAQ/CASE dataset, SVAT obtains much larger IRR than ARIMA/GCN. Again, although the STD of Buy\&Hold's daily returns is the smallest, the IRR curve of Buy\&Hold is inferior to other state-of-the-art stock recommendation models. All of these observations vividly shows that our method effectively reduces the volatility of the investment returns and achieves a good balance between profits and risks.

\subsection{RQ2: Performance Comparison of Financial Crisis Period}

\begin{figure}
\centering

\subfloat[Daily returns of all models. \texttt{STD} denotes the standard deviation of the model profit.]{ 
  \begin{minipage}[t]{\columnwidth}
  \centering
  \includegraphics[width=\columnwidth]{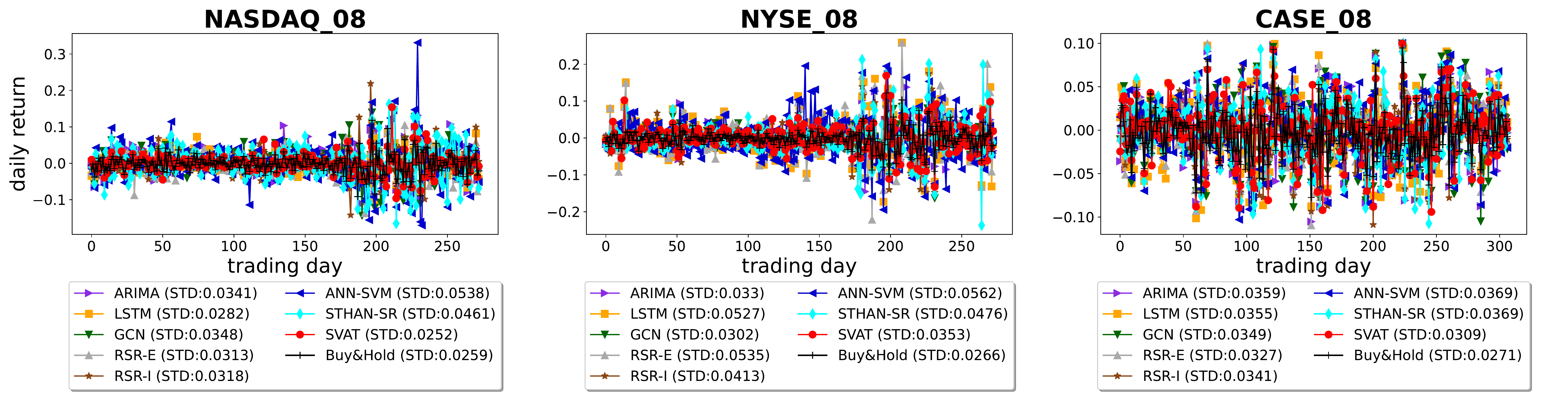}
  \label{fig:daily_returns_crisis}
  \end{minipage}
}

\subfloat[Cumulative investment return ratios (IRR) of all models.]{
  \begin{minipage}[t]{\columnwidth}
  \centering
  \includegraphics[width=\columnwidth]{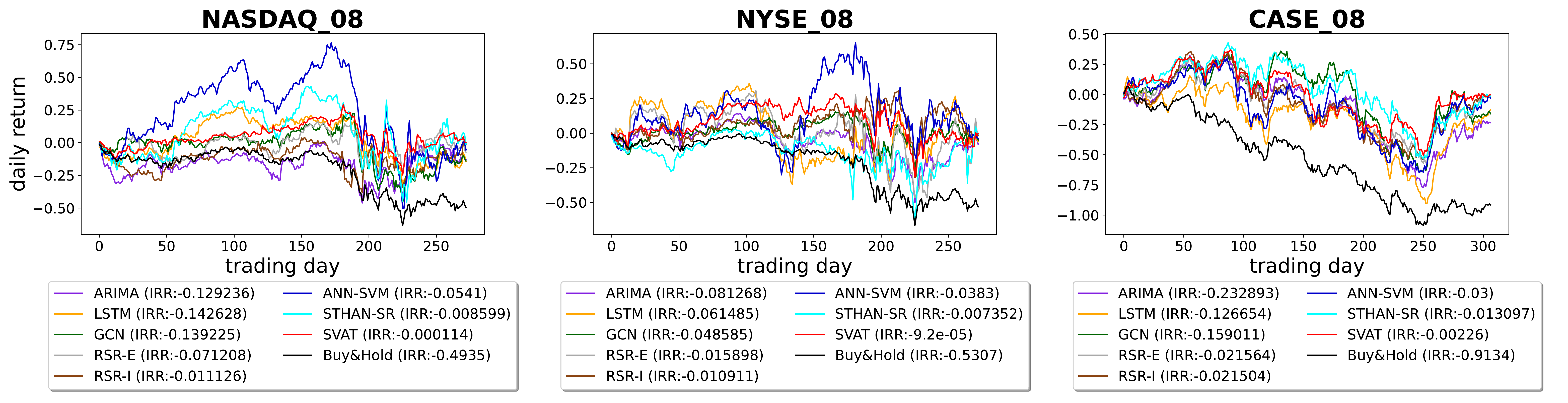}
  \label{fig:cumulative_IRR_crisis}
  \end{minipage}
}
\caption{The curves of daily returns and cumulative investment return ratios of all models backtested on the three crisis datasets. Best viewed in color.}
\Description{Daily and cumulative returns of all models on crisis datasets.}
\label{fig:returns_crisis}
\end{figure}

Table~\ref{tab:result_crisis} shows the experimental results of the buy and hold index and all stock recommendation methods in terms of profitability and risk on the three crisis datasets (NASDAQ\_08, NYSE\_08, CASE\_08), from which we observe that:
\begin{itemize}
  \item Due to the cruelty of the global financial crisis, all methods suffer investment losses (i.e., IRR $<0$) on the three crisis datasets. In particular, although Buy\&Hold achieves the smallest MDD across the three crisis datasets, the IRR and SR of Buy\&Hold are much worse than other stock recommendation methods. This observation further demonstrates that a good stock recommendation model is critical to protecting investors from severe loss during the global financial crisis.
  \item Remarkably, among all stock recommendation methods, our method SVAT achieves the best IRRs and MDDs across all crisis datasets, showcasing that SVAT successfully protects investors from more losses than the other baseline methods. This answers the RQ2 that SVAT can protect investors from risks better than other state-of-the-art stock recommendation models under the extreme circumstance of the global financial crisis.
  \item As for volatility evaluation, SVAT attains the best SR on NYSE\_08 and CASE\_08 datasets but performs worse than the STHAN-SR model on NASDAQ\_08 dataset. Nevertheless, Figure~\ref{fig:daily_returns_crisis} shows that the volatility (i.e., the standard deviation (STD) of the model profit) of SVAT is lower than STHAN-SR on NASDAQ\_08 dataset. This is mainly because according to Equation~(\ref{equ:SR}), the SR metric will decrease as the STD of the model profit decreases when IRR $< 0 < R_f$. Hence, the SR metric might be slightly distorted in the environment of global financial crisis, but Figure~\ref{fig:daily_returns_crisis} demonstrates that our method can still effectively reduce the volatility of the stock recommendation model under extreme circumstances.
  \item Similarly, although the state-of-the-art models RSR-E, RSR-I, ANN-SVM and STHAN-SR obtains better IRR and SR than simple methods ARIMA, LSTM, and GCN on the three crisis datasets, they perform poorly on the MDD metric. Instead, SVAT outperforms all stock recommendation methods on the MDD metric, showing the improvement on anti-risk ability during the global financial crisis period.
\end{itemize}

\begin{table}
    \centering
    \caption{Backtesting results of the three crisis datasets. $\uparrow$ means the larger the better while $\downarrow$ means the smaller the better. Among the results of all stock recommendation methods, the best result is in \textbf{boldface} and the second best result is \underline{underlined}.}
    \label{tab:result_crisis}
    \resizebox{\textwidth}{!}{
    \begin{tabular}{l|ccc|ccc|ccc}
    \toprule[1pt]
    Dataset                       & \multicolumn{3}{c|}{NASDAQ\_08}                                                          & \multicolumn{3}{c|}{NYSE\_08}                                                            & \multicolumn{3}{c}{CASE\_08}                                     \\
    \midrule[.5pt]
    Model                         & IRR$^{\times 10^{-3}}\uparrow$  & SR$^{\times 10^{-2}}\uparrow$  & MDD$\downarrow$       & IRR$^{\times 10^{-3}}\uparrow$  & SR$^{\times 10^{-2}}\uparrow$  & MDD$\downarrow$       & IRR$^{\times 10^{-3}}\uparrow$  & SR$^{\times 10^{-2}}\uparrow$  & MDD$\downarrow$      \\
    \midrule[.5pt]
    Buy\&Hold                     & $-493.53$                       & $-113.76$                      & $9.14\%$              & $-530.75$                       & $-118.83$                      & $9.73\%$              & $-913.38$                       & $-177.13$                      & $7.73\%$             \\
    \midrule[.2pt]
    ARIMA                         & $-129.24$                       & $-24.33$                       & $11.10\%$             & $-81.27$                        & $-16.68$                       & \underline{$14.51\%$} & $-232.89$                       & $-35.75$                       & $10.52\%$            \\
    LSTM                          & $-142.63$                       & $-32.17$                       & \underline{$10.62\%$} & $-61.48$                        & $-8.26$                        & $17.27\%$             & $-126.65$                       & $-20.66$                       & \underline{$10.12\%$}\\
    GCN                           & $-139.23$                       & $-25.54$                       & $14.36\%$             & $-48.59$                        & $-11.96$                       & $16.28\%$             & $-159.01$                       & $-25.76$                       & $10.47\%$            \\
    RSR-E                         & $-71.21$                        & $-15.74$                       & $13.72\%$             & $-15.90$                        & $-3.19$                        & $22.14\%$             & $-21.56$                        & $-5.80$                        & $10.94\%$            \\
    RSR-I                         & $-11.13$                        & $-4.50$                        & $14.23\%$             & $-10.91$                        & $-3.43$                        & $14.74\%$             & $-21.50$                        & $-5.55$                        & $10.87\%$            \\
    ANN-SVM                       & $-54.12$                        & $-7.30$                        & $17.05\%$             & $-38.34$                        & $-5.36$                        & $19.43\%$             & $-30.03$                        & $-6.33$                        & $10.28\%$            \\
    STHAN-SR                      & \underline{$-8.60$}             & $\bm{-2.78}$                   & $16.53\%$             & \underline{$-7.35$}             & \underline{$-2.54$}            & $23.66\%$             & \underline{$-13.10$}            & \underline{$-3.96$}            & $10.73\%$            \\
    SVAT (Ours)                   & $\bm{-0.11}$                    & \underline{${-3.14}$}          & $\bm{9.62\%}$         & $\bm{-0.0916}$                  & $\bm{-2.23}$                   & $\bm{13.23\%}$        & $\bm{-2.26}$                    & $\bm{-2.91}$                   & $\bm{9.39\%}$        \\ 
    \bottomrule[1pt]                                                                                                                                                                                                                                                                                               
    \end{tabular}
    }
\end{table}

Again, we present the curves of daily returns and the cumulative investment return ratios (IRR) of all models backtested on the three crisis datasets in Figure~\ref{fig:returns_crisis}. As expected, the curves of the global financial crisis period are more volatile than that of the normal period shown in Figure~\ref{fig:returns}. However, our method still has better stability than other baseline stock recommendation methods and thus more effectively reduces the risk of losses for investors in the extreme environment of financial crisis. Similarly, although the STD of Buy\&Hold's daily returns is the smallest, the IRR curve of Buy\&Hold is intolerable to most investors.

\subsection{RQ3: Correlations between All Methods}

From Figure~\ref{fig:cumulative_IRR} and Figure~\ref{fig:cumulative_IRR_crisis}, we observe that the cumulative investment returns of different methods behave differently under normal economic environment while exhibit high correlation with each other during the financial crisis period. Such differences inspire us to further investigate the extent to which all methods are correlated in different economic environments. Hence, in this section we aim to evaluate the correlation between different methods by the following two metrics:
\begin{enumerate}
  \item Pearson Correlation Coefficient (PCC): \begin{equation}
  \text{PCC}(m_1, m_2) = \frac{1}{T} \sum_{t} PCC^t(m_1, m_2) = \frac{1}{T} \sum_{t} \frac{\mathbb{E}_{\mathcal{S}}[(\hat{Y}^{m_1}_t - \mathbb{E}_{\mathcal{S}}[\hat{Y}^{m_1}_t]) (\hat{Y}^{m_2}_t - \mathbb{E}_{\mathcal{S}}[\hat{Y}^{m_2}_t])]}{\text{STD}_{\mathcal{S}}[\hat{Y}^{m_1}_t] \cdot \text{STD}_{\mathcal{S}}[\hat{Y}^{m_2}_t]}, \nonumber 
  \end{equation}
  where $m_1, m_2 \in \{\text{ARIMA, LSTM, GCN, RSR-E, RSR-I, ANN-SVM, STHAN-SR, SVAT}\}$ denote any two of all the models, $T$ is the length of trading days, $\hat{Y}^m_t$ is the ranking score predicted by model $m$ on trading day $t$ and $\mathbb{E}_{\mathcal{S}}[\cdot], \text{STD}_{\mathcal{S}}[\cdot]$ denote the expectation and the standard deviation w.r.t the set of all stocks $\mathcal{S}$, respectively. $\text{PCC} \in [-1, 1]$ measures the linear correlation between any two of all the methods where a PCC value close to~$0$ indicates a weak linear relationship between the two methods and vice versa.
  \item Top-$k$ Stocks Difference ($\text{TSD}^k$): \begin{equation}
  \text{TSD}^k(m_1, m_2) = \frac{1}{T} \sum_t \frac{|\mathcal{S}^{m_1}_t - \mathcal{S}^{m_2}_t|}{k}, \nonumber
  \end{equation}
  where $\mathcal{S}^{m}_t \subset \mathcal{S}$ denotes the set of top-$k$ stocks selected by model $m$ on trading day $t$ and thus $|\mathcal{S}^{m_1}_t - \mathcal{S}^{m_2}_t|$ is the number of different stocks among the top-$k$ stocks selected by models $m_1$ and $m_2$. $\text{TSD}^k \in [0, 1]$ measures the extent of how any two of all the methods recommend stocks different from each other, with a $\text{TSD}^k$ value close to~$1$ indicating a weak stock-selection relationship between the two methods and vice versa.
\end{enumerate}

\begin{figure}
\centering

\subfloat[Model PCCs on the three normal datasets.]{ 
  \begin{minipage}[t]{\columnwidth}
  \centering
  \includegraphics[width=\columnwidth]{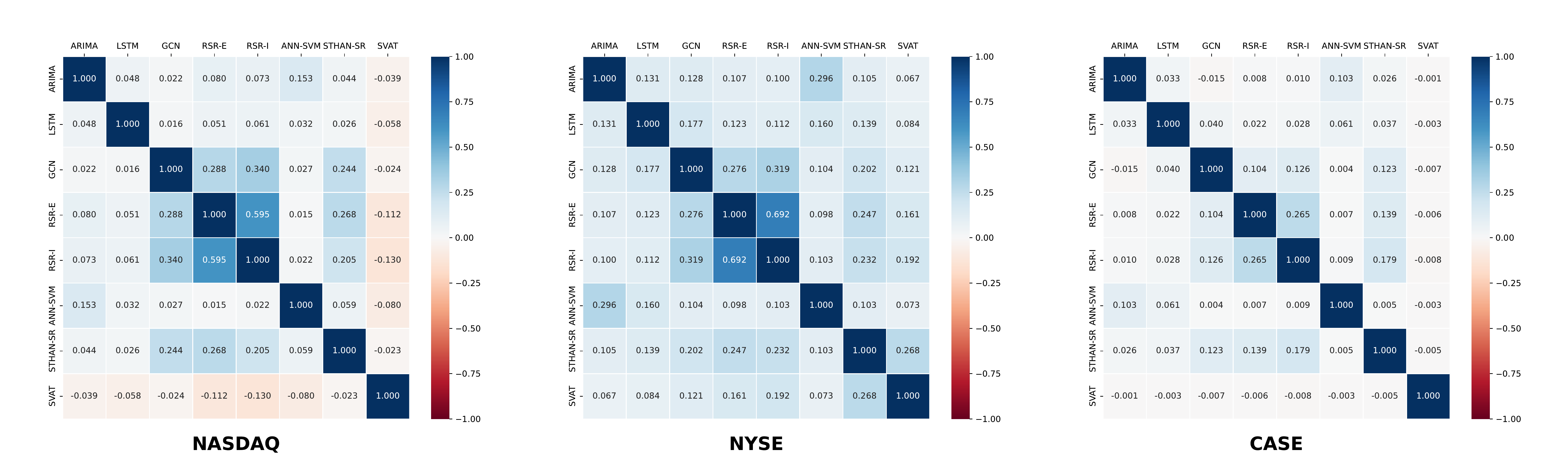}
  \label{fig:normal_PCC}
  \end{minipage}
}

\subfloat[Model PCCs on the three crisis datasets.]{
  \begin{minipage}[t]{\columnwidth}
  \centering
  \includegraphics[width=\columnwidth]{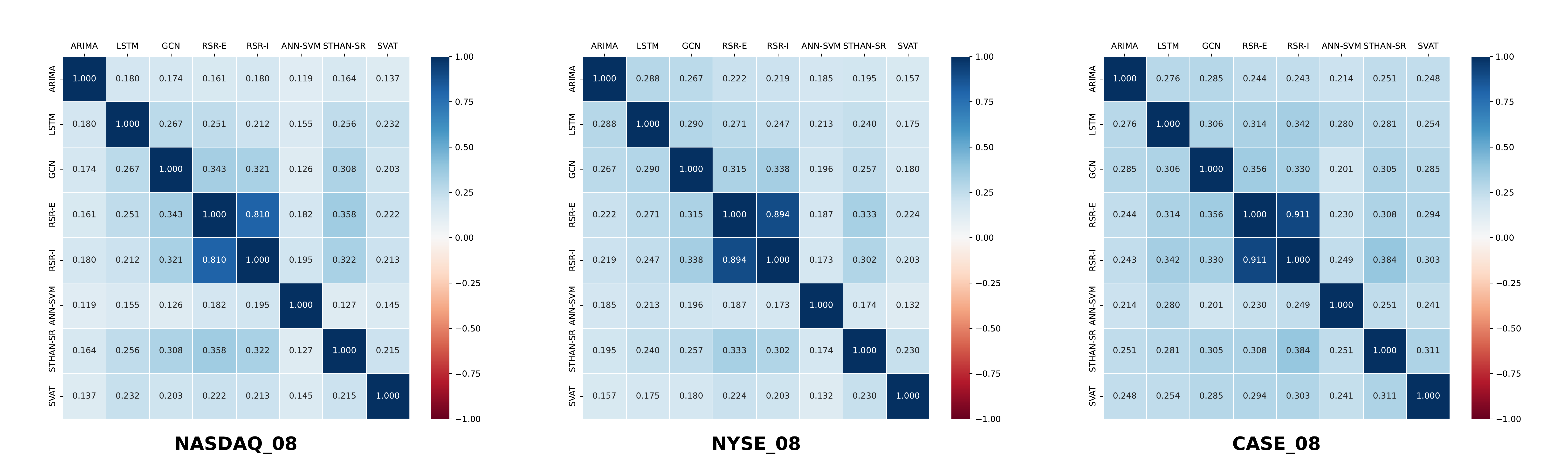}
  \label{fig:crisis_PCC}
  \end{minipage}
}
\caption{The Pearson Correlation Coefficient (PCC) between any two of all the methods on all datasets.}
\Description{PCCs of all models.}
\label{fig:PCC}
\end{figure}

\begin{figure}
\centering

\subfloat[Model $\text{TSD}^5$ on the three normal datasets.]{ 
  \begin{minipage}[t]{\columnwidth}
  \centering
  \includegraphics[width=\columnwidth]{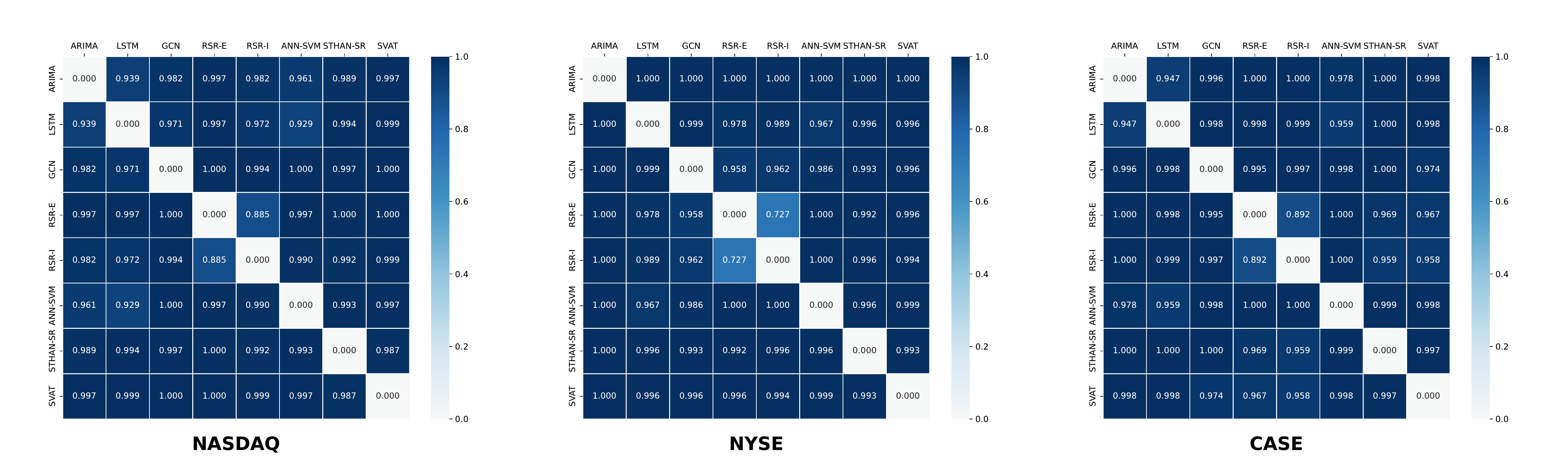}
  \label{fig:normal_TSD}
  \end{minipage}
}

\subfloat[Model $\text{TSD}^5$ on the three crisis datasets.]{
  \begin{minipage}[t]{\columnwidth}
  \centering
  \includegraphics[width=\columnwidth]{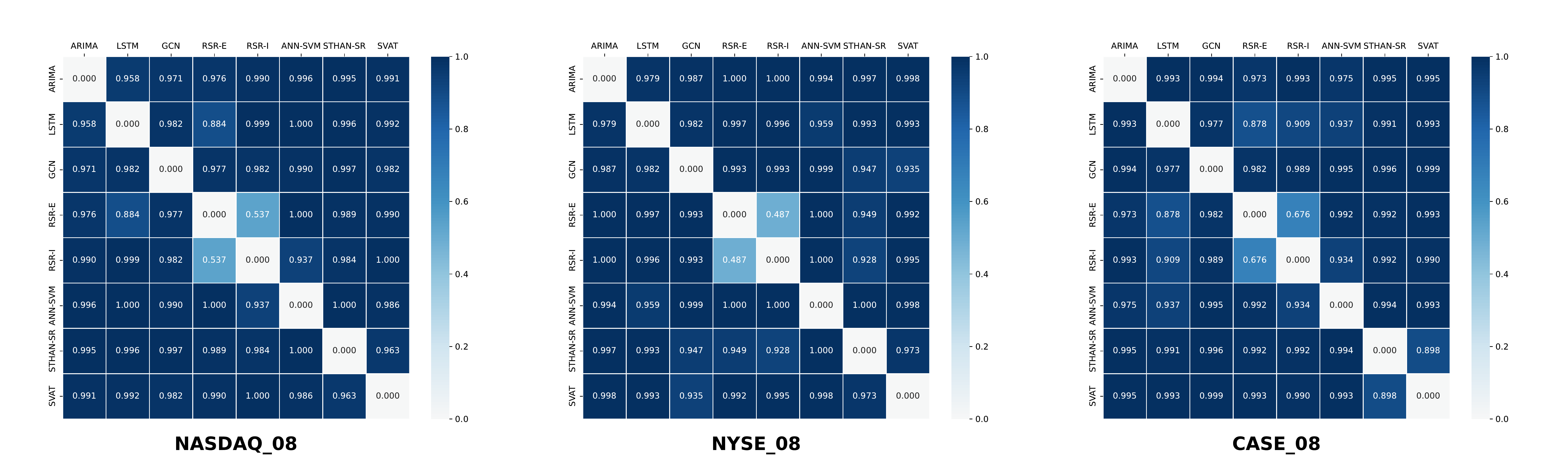}
  \label{fig:crisis_TSD}
  \end{minipage}
}
\caption{The Top-$5$ Stocks Difference ($\text{TSD}^5$) between any two of all the methods on all datasets.}
\Description{TSDs of all models.}
\label{fig:TSD}
\end{figure}

\begin{figure}
  \centering
  \includegraphics[width=\linewidth]{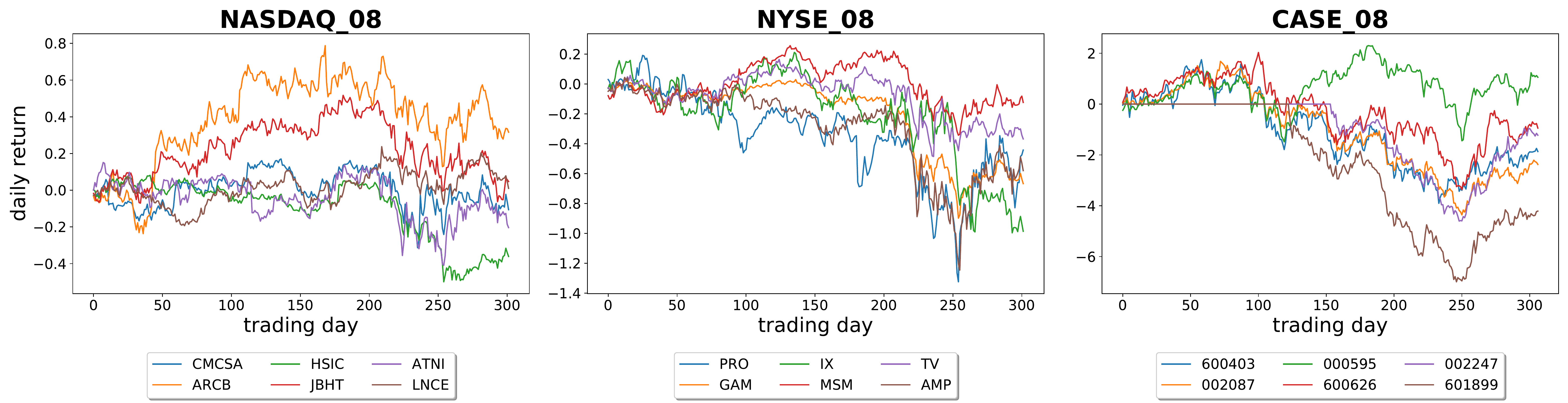}
  \caption{Cumulative investment return ratios (IRR) of six randomly selected stocks in each crisis dataset. Different stocks experienced similar drawdowns during the global financial crisis.}
  \label{fig:crisis_return}
  \Description{Stock returns on the crisis datasets.}
\end{figure}

Figure~\ref{fig:PCC} and Figure~\ref{fig:TSD} show the PCC and $\text{TSD}^5$ between any two of all the methods, from which we can obtain the following observations about model correlation:
\begin{itemize}
  \item According to the PCC values in Figure~\ref{fig:PCC}, the four models including GCN, RSR-E, RSR-I, and STHAN-SR are more correlated to each other than other models, mainly because that all of them are graph-based learning methods. Although our SVAT method employs STHAN-SR as the backbone model, it presents some different correlations compared to STHAN-SR and other graph-based models on the three normal datasets. Nevertheless, Figure~\ref{fig:TSD} shows that the $\text{TSD}^5$ values between any two of all the methods are close to $0.9$ or above, except the two similar models RSR-E and RSR-I. The high $\text{TSD}^5$ values demonstrate that most methods have a weak stock-selection relationship and tend to independently recommend stocks different from other methods. 
  \item Particularly on the three normal datasets, our SVAT method exhibits weak negative correlations with other methods on NASDAQ and CASE datasets, while maintaining weak positive correlations on NYSE dataset. Recalling the fact that NASDAQ and CASE stock markets are more volatile than NYSE stock market~\cite{STHAN-SR}, we infer that the proposed split variational adversarial training framework is better at capturing different trading signals in more volatile markets.
  \item Paradoxically, both the cumulative investment returns in Figure~\ref{fig:cumulative_IRR_crisis} and the PCC values in Figure~\ref{fig:crisis_PCC} show that all methods have relatively high correlations on the three crisis datasets, but the high $\text{TSD}^5$ values in Figure~\ref{fig:crisis_TSD} reveals that most methods have weak stock-selection relationships. We speculate that under the extreme situation of the global financial crisis, most stocks suffered from similar drawdowns. Figure~\ref{fig:crisis_return} presents the cumulative return curves of six stocks (randomly selected) of the three crisis datasets during the global financial crisis, from which we observe that all these stocks experienced similar drawdowns and their cumulative investment returns are highly correlated. Therefore, although different methods will recommend different stocks to investors, they produce highly correlated cumulative investment returns during the global financial crisis.
\end{itemize}

\subsection{RQ4: Effectiveness of Model Design}

\subsubsection{Effects of SVAT on Other Baselines}

We have demonstrated that incorporating SVAT with the backbone model STHAN-SR can achieve the best results against other state-of-the-art baselines. In this section, we further evaluate the overall performance of the SVAT method by inspecting whether SVAT could also improve the performance of other baselines. Hence, we combine the SVAT method with LSTM, GCN, RSR-E, RSR-I, ANN-SVM models for stock recommendation, respectively, and compare their results with original models.

Table~\ref{tab:SVAT_other_effect} shows the comparison results between other baselines and their SVAT-variants on the normal and the crisis datasets. Among the total 90 comparison cases, the SVAT-variants score 85 better results. This further demonstrates that SVAT is a general learning framework that can be incorporated with various stock recommendation models to enhance their risk awareness.

\begin{table}
    \centering
    \caption{Comparison between other baselines and their SVAT-variants on the normal and the crisis datasets. Better results are in \textbf{boldface}.}
    \label{tab:SVAT_other_effect}
    \resizebox{\textwidth}{!}{
    \begin{tabular}{l|ccc|ccc|ccc|c}
    \toprule[1pt]
    Dataset                 & \multicolumn{3}{c|}{NASDAQ}                             & \multicolumn{3}{c|}{NYSE}                               & \multicolumn{3}{c|}{CASE}                          & Better Results  \\
    \midrule[.5pt]
    Model                   & IRR$\uparrow$  & SR$\uparrow$   & MDD$\downarrow$       & IRR$\uparrow$  & SR$\uparrow$   & MDD$\downarrow$       & IRR$\uparrow$  & SR$\uparrow$   & MDD$\downarrow$  & Count           \\
    \midrule[.5pt]
    LSTM                    & $0.22$         & $0.95$         & $7.37\%$              & $0.12$         & $0.79$         & $5.72\%$              & $0.35$         & $0.53$         & $7.75\%$         & $0$              \\
    SVAT+LSTM               & $\bm{0.37}$    & $\bm{1.67}$    & $\bm{3.64\%}$         & $\bm{0.25}$    & $\bm{0.92}$    & $\bm{5.05\%}$         & $\bm{0.37}$    & $\bm{0.95}$    & $\bm{6.92\%}$    & $9$              \\
    \midrule[.2pt]
    GCN                     & $0.13$         & $0.46$         & $7.91\%$              & $0.16$         & $0.72$         & $6.20\%$              & $0.43$         & $1.03$         & $7.33\%$         & $0$              \\
    SVAT+GCN                & $\bm{0.24}$    & $\bm{1.26}$    & $\bm{4.19\%}$         & $\bm{0.24}$    & $\bm{0.83}$    & $\bm{5.99\%}$         & $\bm{0.53}$    & $\bm{1.30}$    & $\bm{6.41\%}$    & $9$              \\ 
    \midrule[.2pt]
    RSR-E                   & $\bm{0.26}$    & $1.12$         & $7.35\%$              & $0.20$         & $0.88$         & $5.86\%$              & $\bm{0.56}$    & $0.96$         & $5.95\%$         & $2$              \\
    SVAT+RSR-E              & $0.21$         & $\bm{1.16}$    & $\bm{3.55\%}$         & $\bm{0.25}$    & $\bm{1.03}$    & $\bm{4.97\%}$         & $0.55$         & $\bm{1.26}$    & $\bm{4.85\%}$    & $7$              \\ 
    \midrule[.2pt]
    RSR-I                   & $\bm{0.39}$    & $1.34$         & $6.75\%$              & $0.21$         & $0.95$         & $6.34\%$              & $0.58$         & $1.04$         & $7.95\%$         & $1$              \\
    SVAT+RSR-I              & $0.36$         & $\bm{1.79}$    & $\bm{5.72\%}$         & $\bm{0.30}$    & $\bm{1.11}$    & $\bm{5.13\%}$         & $\bm{0.60}$    & $\bm{1.36}$    & $\bm{6.72\%}$    & $8$              \\ 
    \midrule[.2pt]
    ANN-SVM                 & $0.32$         & $1.28$         & $5.73\%$              & $0.33$         & $1.14$         & $8.59\%$              & $0.26$         & $0.43$         & $5.97\%$         & $0$              \\
    SVAT+ANN-SVM            & $\bm{0.48}$    & $\bm{1.84}$    & $\bm{4.51\%}$         & $\bm{0.41}$    & $\bm{1.17}$    & $\bm{8.40\%}$         & $\bm{0.42}$    & $\bm{1.06}$    & $\bm{4.74\%}$    & $9$              \\ 
    \midrule[.5pt]
    \midrule[.5pt]
    Dataset                 & \multicolumn{3}{c|}{NASDAQ\_08}                                                          & \multicolumn{3}{c|}{NYSE\_08}                                                            & \multicolumn{3}{c|}{CASE\_08}                                                         & Better Results   \\
    \midrule[.5pt]
    Model                   & IRR$^{\times 10^{-3}}\uparrow$  & SR$^{\times 10^{-2}}\uparrow$  & MDD$\downarrow$       & IRR$^{\times 10^{-3}}\uparrow$  & SR$^{\times 10^{-2}}\uparrow$  & MDD$\downarrow$       & IRR$^{\times 10^{-3}}\uparrow$  & SR$^{\times 10^{-2}}\uparrow$  & MDD$\downarrow$    & Count            \\
    \midrule[.5pt]
    LSTM                    & $-142.63$                       & $-32.17$                       & $10.62\%$             & $-61.48$                        & $-8.26$                        & $17.27\%$             & $-126.65$                       & $-20.66$                       & $10.12\%$          & $0$                 \\
    SVAT+LSTM               & $\bm{-79.31}$                   & $\bm{-14.84}$                  & $\bm{9.85\%}$         & $\bm{-53.00}$                   & $\bm{-7.93}$                   & $\bm{14.91\%}$        & $\bm{-64.41}$                   & $\bm{-11.07}$                  & $\bm{9.86\%}$      & $9$                 \\
    \midrule[.2pt]
    GCN                     & $-139.23$                       & $-25.54$                       & $14.36\%$             & $-48.59$                        & $-11.96$                       & $16.28\%$             & $-159.01$                       & $-25.76$                       & $10.47\%$          & $0$                 \\
    SVAT+GCN                & $\bm{-68.44}$                   & $\bm{-8.23}$                   & $\bm{13.70\%}$        & $\bm{-34.21}$                   & $\bm{-4.44}$                   & $\bm{14.84\%}$        & $\bm{-71.21}$                   & $\bm{-12.86}$                  & $\bm{10.34\%}$     & $9$                 \\ 
    \midrule[.2pt]
    RSR-E                   & $-71.21$                        & $-15.74$                       & $13.72\%$             & $-15.90$                        & $\bm{-3.19}$                   & $22.14\%$             & $-21.56$                        & $-5.80$                        & $10.94\%$          & $1$                 \\
    SVAT+RSR-E              & $\bm{-59.00}$                   & $\bm{-12.37}$                  & $\bm{9.26\%}$         & $\bm{-10.99}$                   & $-3.60$                        & $\bm{14.22\%}$        & $\bm{-4.83}$                    & $\bm{-2.64}$                   & $\bm{9.98\%}$      & $8$                 \\ 
    \midrule[.2pt]
    RSR-I                   & $-11.13$                        & $-4.50$                        & $14.23\%$             & $-10.91$                        & $-3.43$                        & $14.74\%$             & $-21.50$                        & $-5.55$                        & $10.87\%$          & $0$                 \\
    SVAT+RSR-I              & $\bm{-3.20}$                    & $\bm{-3.63}$                   & $\bm{8.60\%}$         & $\bm{-3.03}$                    & $\bm{-2.64}$                   & $\bm{14.00\%}$        & $\bm{-4.17}$                    & $\bm{-2.61}$                   & $\bm{10.01\%}$     & $9$                 \\ 
    \midrule[.2pt]
    ANN-SVM                 & $-54.12$                        & $\bm{-7.30}$                   & $17.05\%$             & $-38.34$                        & $-5.36$                        & $19.43\%$             & $-30.03$                        & $-6.33$                        & $10.28\%$          & $1$                 \\
    SVAT+ANN-SVM            & $\bm{-35.13}$                   & $-8.05$                        & $\bm{12.27\%}$        & $\bm{-26.12}$                   & $\bm{-4.38}$                   & $\bm{17.96\%}$        & $\bm{-17.41}$                   & $\bm{-4.89}$                   & $\bm{9.98\%}$      & $8$                  \\ 
    \bottomrule[1pt]
    \end{tabular}}
\end{table}

\subsubsection{Ablation Study of Key Components}

Our method consists of two key components, the split adversarial training mechanism (Equation~(\ref{equ:adv_loss})) and the variational perturbation generator (VPG), as shown in Figure~\ref{fig:SVAT_VPG}. In this experiment, we aim to evaluate whether these two components are necessary to reduce stock recommendation risks. Therefore, to show the effectiveness of each component, we compare SVAT with two variants:
\begin{itemize}
    \item \textbf{SVATw/oS}: We change the adversarial loss in Equation~(\ref{equ:adv_loss}) to be $\mathcal{L}^{adv} = \sum_{i=1}^N \mathcal{L}^{adv}_i$ without weighting by stocks' return ratios, which reduces SVAT to conventional adversarial training without the split effect.
    \item \textbf{SVATw/oV}: We remove the VPG from SVAT and only employ Equation~(\ref{equ:post_delta}) to generate the perturbation for each stock example.
\end{itemize}

Table~\ref{tab:ablation} presents the results of comparison on the normal and the crisis datasets. We can observe that SVAT attains higher IRR and SR than all variants on most datasets. However, the MDD of SVAT is relatively inferior to its variants on NASDAQ, NYSE, and NYSE\_08 datasets. We postulate the reason is that either the split effect or the VPG mainly works to reduce the risk of maximum daily loss without significant profit improvement, while combining them into the complete SVAT framework can achieve a better balance between profits and risks. As a result, both the split adversarial training mechanism and the variational perturbation generator are necessary to construct the complete SVAT architecture for better stock recommendation.

\begin{table}
    \centering
    \caption{Comparison between SVAT and other variants on the normal and the crisis datasets. The best result is in \textbf{boldface}.}
    \label{tab:ablation}
    \resizebox{\textwidth}{!}{
    \begin{tabular}{l|ccc|ccc|ccc}
    \toprule[1pt]
    Dataset                 & \multicolumn{3}{c|}{NASDAQ}                                       & \multicolumn{3}{c|}{NYSE}                                         & \multicolumn{3}{c}{CASE}                                         \\
    \midrule[.5pt]
    Model                   & IRR$\uparrow$       & SR$\uparrow$        & MDD$\downarrow$       & IRR$\uparrow$       & SR$\uparrow$        & MDD$\downarrow$       & IRR$\uparrow$       & SR$\uparrow$        & MDD$\downarrow$      \\
    \midrule[.5pt]
    SVATw/oS                & $0.47$              & $2.84$              & $3.68\%$              & $0.12$              & $1.26$              & $\bm{1.77\%}$         & $0.69$              & $1.26$              & $6.69\%$             \\
    SVATw/oV                & $0.42$              & $2.20$              & $\bm{3.19\%}$         & $0.21$              & $1.75$              & $1.88\%$              & $0.68$              & $1.02$              & $6.56\%$             \\
    SVAT                    & $\bm{0.59}$         & $\bm{3.10}$         & $3.29\%$              & $\bm{0.38}$         & $\bm{2.61}$         & $3.13\%$              & $\bm{0.79}$         & $\bm{1.50}$         & $\bm{5.88\%}$        \\ 
    \midrule[.5pt]
    \midrule[.5pt]
    Dataset                 & \multicolumn{3}{c|}{NASDAQ\_08}                                                          & \multicolumn{3}{c|}{NYSE\_08}                                                            & \multicolumn{3}{c}{CASE\_08}                                     \\
    \midrule[.5pt]
    Model                   & IRR$^{\times 10^{-3}}\uparrow$  & SR$^{\times 10^{-2}}\uparrow$  & MDD$\downarrow$       & IRR$^{\times 10^{-3}}\uparrow$  & SR$^{\times 10^{-2}}\uparrow$  & MDD$\downarrow$       & IRR$^{\times 10^{-3}}\uparrow$  & SR$^{\times 10^{-2}}\uparrow$  & MDD$\downarrow$      \\
    \midrule[.5pt]
    SVATw/oS                & $-2.46$                         & $-2.41$                        & $11.76\%$             & $-89.13$                        & $-15.35$                       & $16.68\%$             & $-14.29$                        & $-4.42$                        & $9.97\%$              \\
    SVATw/oV                & $-0.74$                         & $\bm{-2.39}$                   & $15.76\%$             & $-14.04$                        & $-5.66$                        & $\bm{8.35\%}$         & $-15.62$                        & $-4.43$                        & $9.97\%$              \\
    SVAT                    & $\bm{-0.11}$                    & $-3.14$                        & $\bm{9.62\%}$         & $\bm{-0.0916}$                  & $\bm{-2.23}$                   & $13.23\%$             & $\bm{-2.26}$                    & $\bm{-2.91}$                   & $\bm{9.39\%}$         \\ 
    \bottomrule[1pt]
    \end{tabular}}
\end{table}

\subsection{RQ5: Parameter Sensitivity Analysis}

\subsubsection{Performance under Different Backtesting Strategies}

\begin{figure}
\centering

\subfloat[Comparison on different backtesting strategies (top $1$-$10$) with regard to the sharp ratio (SR) on the three normal datasets.]{ 
  \begin{minipage}[t]{\columnwidth}
  \centering
  \includegraphics[width=\columnwidth]{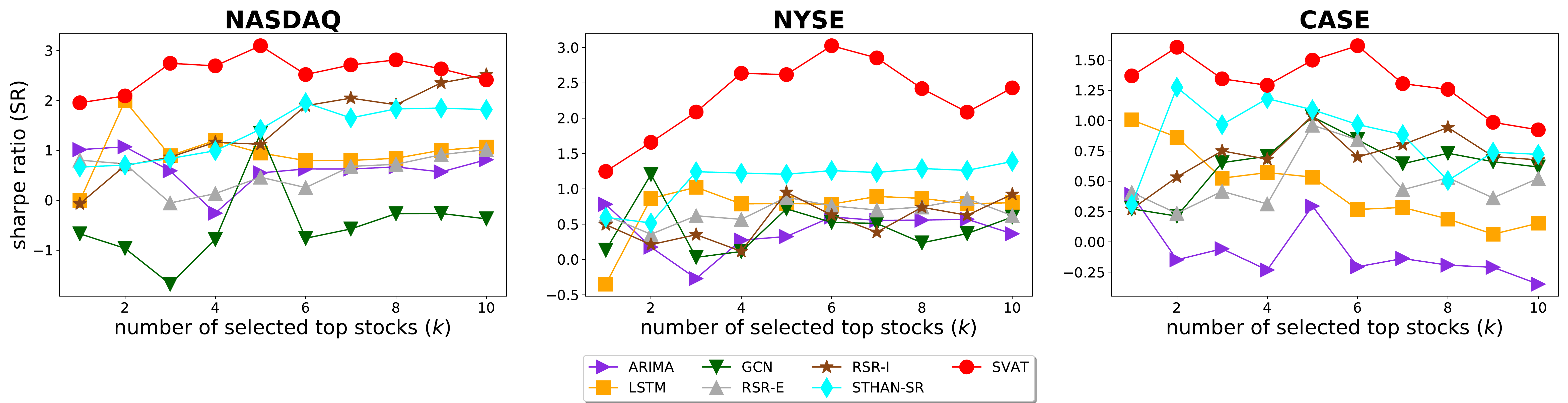}
  \label{fig:normal_para_sen}
  \end{minipage}
}

\subfloat[Comparison on different backtesting strategies (top $1$-$10$) with regard to the sharp ratio (SR) on the three crisis datasets.]{
  \begin{minipage}[t]{\columnwidth}
  \centering
  \includegraphics[width=\columnwidth]{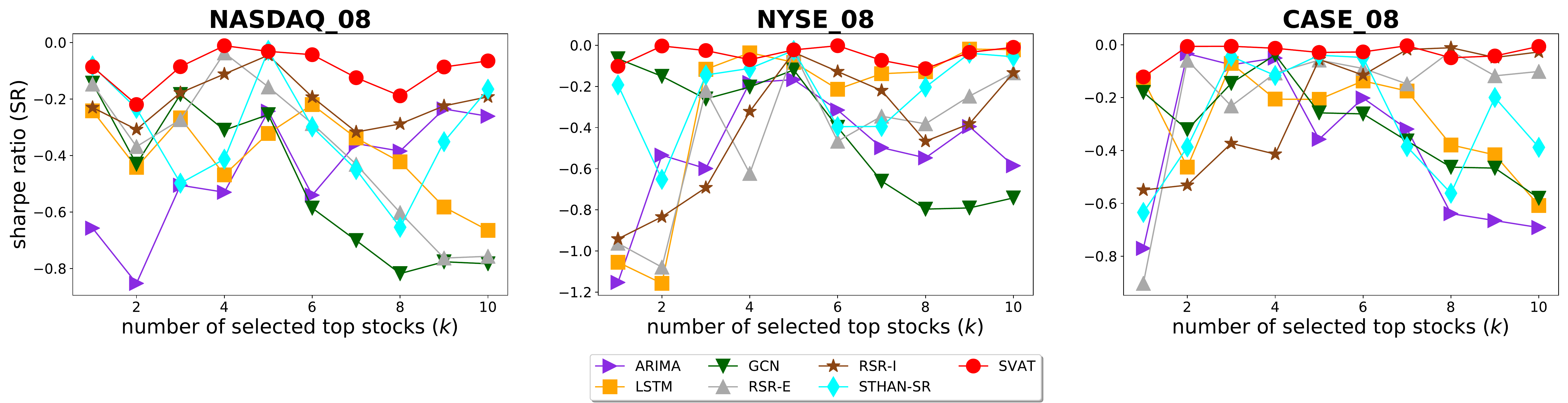}
  \label{fig:crisis_para_sen}
  \end{minipage}
}
\caption{Comparison of model sensitivity to the strategy parameter $k$.}
\Description{Sensitivity of the parameter k on the normal and the crisis datasets.}
\label{fig:para_sen}
\end{figure}

In the scenario of stock recommendation, the core strategy is to select top-$k$ stocks for investment, where different values of $k$ could derive diverse investment strategies. In previous experiments, we follow the previous work~\cite{RSR,STHAN-SR,ALSP_TF} and generally set $k=5$ in order to make consistent comparison with other baseline methods. In practice, however, investors are free to choose different values of $k$ to develop more profitable investment strategies. Therefore, it is necessary to investigate the performance of our proposed method against other baselines under different backtesting strategies. In this section, we conduct backtesting on all methods with $k \in \{1,2,3,4,5,6,7,8,9,10\}$, corresponding to strategies of buying stocks with top-$1$, top-$2$, ..., top-$10$ highest expected returns, respectively. Figure~\ref{fig:para_sen} presents the sharp ratios (SRs) of each model backtesting on the normal and the crisis datasets with different strategies, from which we have the following observations:
\begin{itemize}
  \item Clearly, our method SVAT achieves the best results under different backtesting strategies on almost all datasets, demonstrating the strong adaptability of SVAT such that it is applicable to a wide variety of strategies. 
  \item While most baseline methods stably attain good performance across different strategies on the normal datasets, they experience high volatility under different strategies on the crisis datasets. This observation exposes that under the extreme circumstances of the global financial crisis, existing stock recommendation baselines are prone to be vulnerable and their performance is highly dependent on the investment strategy. Instead, SVAT consistently shows great robustness across different strategies even on the crisis datasets, demonstrating the advantage of incorporating our split variational adversarial training framework for risk-aware stock recommendation.
\end{itemize}

\subsubsection{Impact of the Adversarial Hyperparameter}

\begin{figure}
  \centering
  \includegraphics[width=0.76\linewidth]{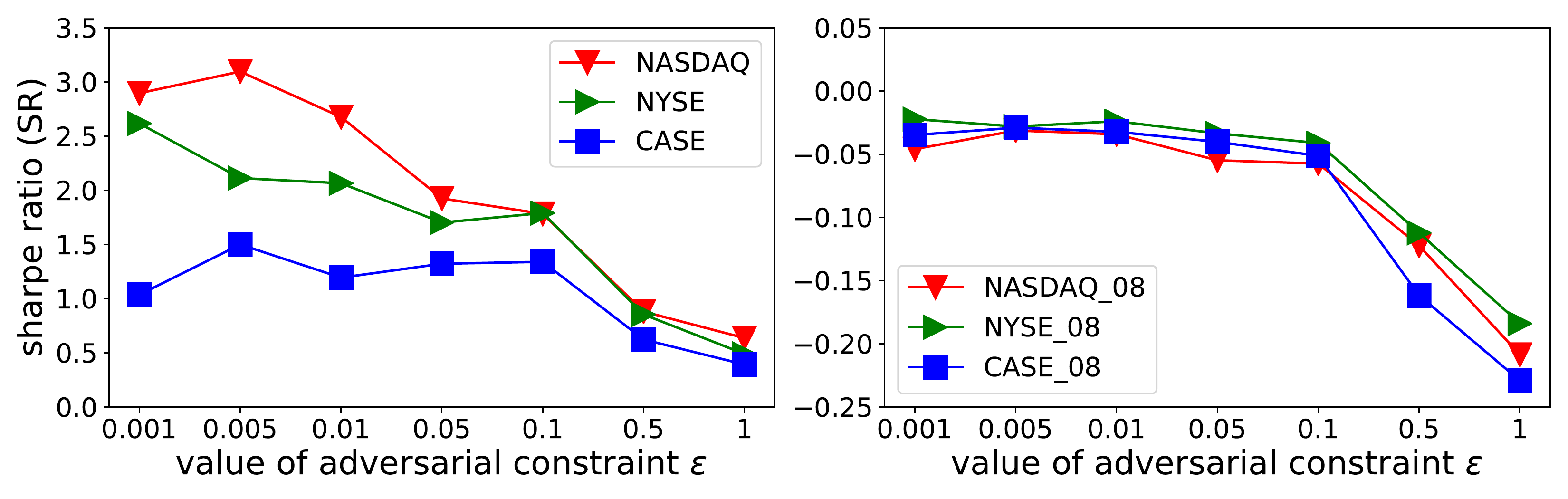}
  \caption{Sensitivity of our method SVAT with regard to the adversarial hyperparameter $\epsilon$.}
  \label{fig:epsilon_para_sen}
  \Description{Sensitivity of the parameter epsilon on the normal and the crisis datasets.}
\end{figure}

For adversarial learning methods, the adversarial hyperparameter $\epsilon$ in Equation~(\ref{equ:post_delta}) plays an important role in the model performance~\cite{adv_epsilon}. In this section, we investigate how the performance of our method SVAT varies with different $\epsilon \in \{0.001, 0.005, 0.01, 0.05, 0.1, 0.5, 1\}$. Figure~\ref{fig:epsilon_para_sen} presents the sharp ratio (SR) of SVAT backtesting on all datasets with different values of $\epsilon$. Obviously, the performance of SVAT drops significantly when $\epsilon > 0.1$, since a large $\epsilon$ may produce excessive perturbations that are harmful for model training. On the contrary, SVAT shows good robustness and stability when $\epsilon$ is controlled within a reasonable range $[0.001, 0.05]$.

\subsection{RQ6: Risk Quantification by Ranking Entropy}

\begin{figure}[t]
\centering

\subfloat[NASDAQ dataset.]{ 
  \begin{minipage}[t]{\columnwidth}
  \centering
  \includegraphics[width=0.72\columnwidth]{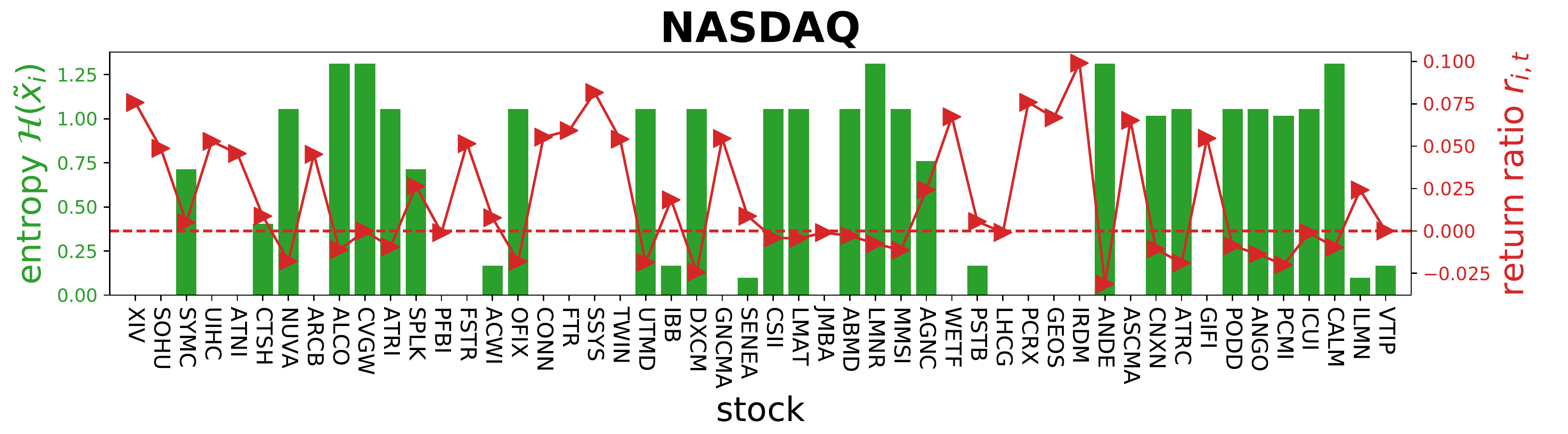}
  \label{fig:NASDAQ_ent}
  \end{minipage}
}

\subfloat[NYSE dataset.]{
  \begin{minipage}[t]{\columnwidth}
  \centering
  \includegraphics[width=0.72\columnwidth]{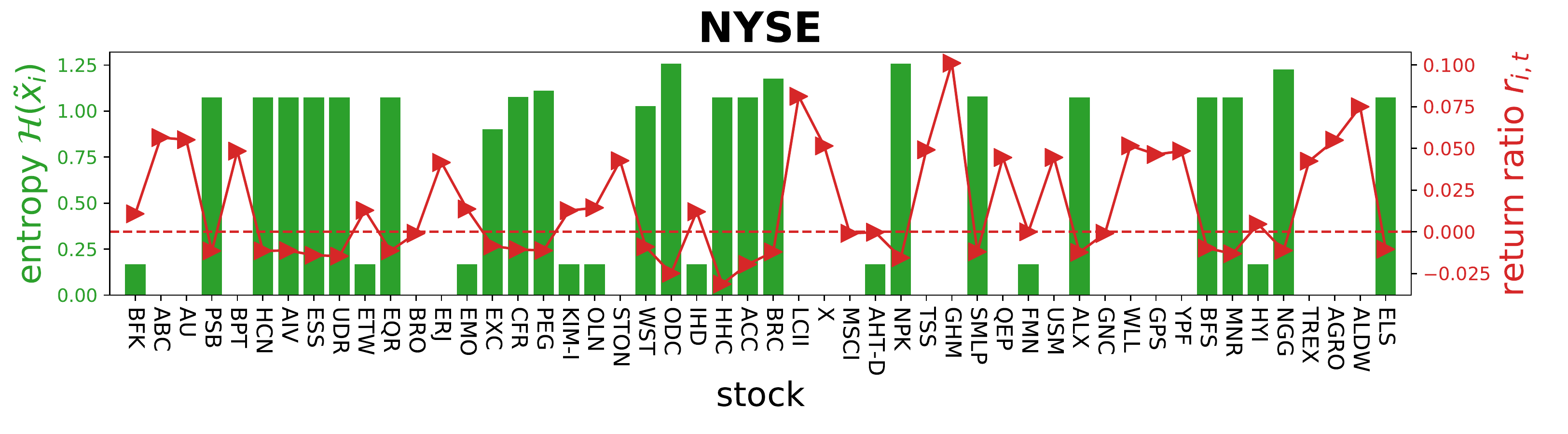}
  \label{fig:NYSE_ent}
  \end{minipage}
}

\subfloat[CASE dataset.]{
  \begin{minipage}[t]{0.72\columnwidth}
  \centering
  \includegraphics[width=\columnwidth]{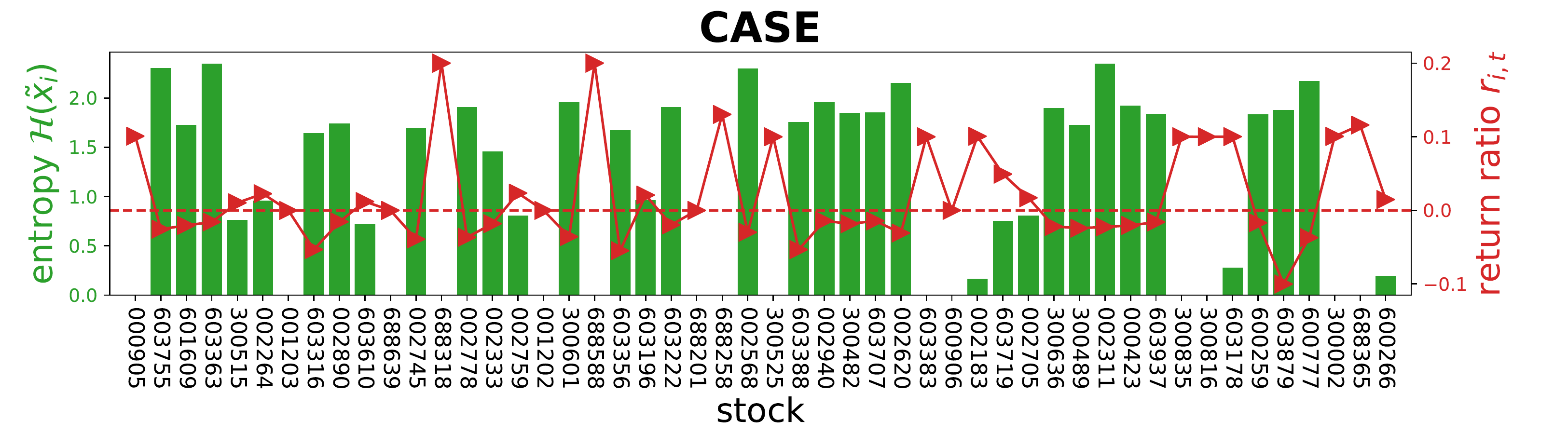}
  \label{fig:CASE_ent}
  \end{minipage}
}
\caption{Relation between ranking entropy and return ratio.}
\Description{Relation between ranking entropy and return ratio.}
\label{fig:rank_entropy}
\end{figure}

In this section, we aim to investigate how the adversarial perturbation of SVAT works in reducing stock recommendation risks and provide some insights about risk quantification for investors. As discussed in Section~\ref{subsec:model_overview}, by sampling multiple adversarial perturbations $\bm{\delta}_i^1, \bm{\delta}_i^2, ..., \bm{\delta}_i^M$ from the perturbation distribution $p(\bm{\delta}_i|\tilde{\mathbf{x}}_i)$ learned by the variational perturbation generator, one can compute the ranking entropy $\mathcal{H}(\tilde{\mathbf{x}}_i)$ for each stock $s_i$ using Equation~(\ref{equ:rank_entropy}). We postulate that the relationship between ranking entropies and stock returns could indirectly reveal the truth about risk reduction. Accordingly, we sample $M=50$ perturbations for each stock example of the testing datasets and inspect the interactions between their ranking entropies and investment returns. Figure~\ref{fig:rank_entropy} shows an example of the relation between $\mathcal{H}(\tilde{\mathbf{x}}_i)$ and the return ratio $r_{i,t}$, where we randomly select $50$ stocks from NASDAQ, NYSE, and CASE datasets for illustration. Overall, the ranking entropy and the return ratio present an inverse relation, where:
\begin{itemize}
  \item For NASDAQ and NYSE datasets, most stocks with $\mathcal{H}(\tilde{\mathbf{x}}_i) < 0.25$ earn profits ($r_{i,t} > 0$) and most stocks with $\mathcal{H}(\tilde{\mathbf{x}}_i) > 0.75$ indicate risks ($r_{i,t} < 0$).
  \item For CASE datasets, most stocks with $\mathcal{H}(\tilde{\mathbf{x}}_i) < 1.0$ earn profits ($r_{i,t} > 0$) and most stocks with $\mathcal{H}(\tilde{\mathbf{x}}_i) > 1.5$ indicate risks ($r_{i,t} < 0$).
\end{itemize}
Since all perturbations are sampled from the distribution $p(\bm{\delta}_i|\tilde{\mathbf{x}}_i)$, the ranking entropy $\mathcal{H}(\tilde{\mathbf{x}}_i)$ actually reflects the uncertainty of $p(\bm{\delta}_i|\tilde{\mathbf{x}}_i)$, with \emph{higher entropy indicating larger uncertainty and thus higher risk of stock $s_i$}. In this case, Figure~\ref{fig:rank_entropy} shows that by increasing the ranking uncertainties of risky stocks with negative returns, the perturbations produced by SVAT probably help the model better recognize those risky stocks and thus avoid recommending such stocks, which reduces the possibility of potential losses and lowers the risk of stock recommendation. Also, the ranking entropy provides an effective tool for investors to roughly evaluate the risk of each stock before making investment decisions, demonstrating the additional advantage of risk interpretability brought by SVAT.

\section{Related Work}
\label{sec:related_work}

Our work is directly related to the recent work on \emph{stock prediction}, \emph{risk management}, \emph{adversarial training} and \emph{variational autoencoder}.

\subsection{Stock Prediction}

Researchers have been studying the stock market for decades, developing various stock prediction methods to pursue excess investment profits~\cite{stock_survey}. As we mentioned in Section~\ref{sec:intro}, recent work on stock prediction can be separated into three categories: \emph{stock price regression}, \emph{stock trend classification}, and \emph{stock recommendation}~\cite{stock_survey_02}. Stock price regression formulates stock prediction as a pure time series forecasting problem and predict the future stock prices/returns by learning from historical stock time series. Traditional econometrics~\cite{econometric} employs simple linear models such as VAR and ARIMA~\cite{stock_TS} to perform stock price regression prediction, while contemporary methods leverage advanced learning architectures such as SVM~\cite{text_stock_pred}, Tensor Aggregation~\cite{tensor_stock_pred}, RNN~\cite{DARNN,MLCNN}, GAN~\cite{stockGAN_01,Zibin_03}, and Transformer~\cite{LogTrans} to better capture non-linear relationships in the stock time series. On the other hand, stock trend classification treats stock prediction as a binary up/down classification problem and develop efficient classifiers to perform stock movement prediction. Advanced classification models include StockNet~\cite{stocknet}, Adv-ALSTM~\cite{adv_LSTM}, STLAT~\cite{Zibin_02}, etc. Nonetheless, both stock price regression and stock trend classification have a significant drawback that they are not directly optimized towards the target of investment (i.e., profit maximization), which limits their practicality~\cite{STHAN-SR}. To overcome this drawback, reinforcement learning methods~\cite{RL_trading_01,RL_trading_02,RL_trading_03} have been proposed to improve model profits by capturing trading signals in a dynamic prediction. And stock recommendation proposes to rank stocks with return ratios based on the comparison among multiple stocks~\cite{TRAN}. In this case, models are trained to select top-$k$ stocks with maximum expected profits so as to ensure their consistency with the investment target. Recent stock recommendation models combine both the RNN and the GNN architectures to learn the relationships between multiple stocks~\cite{RSR,TRAN,STHAN-SR,ALSP_TF}.

Although various stock recommendation methods have achieved promising results, literature on reducing stock recommendation risks with adversarial training is still scarce. In this work, we aim to tackle risk concerns of stock recommendation and demonstrate a new way to reduce investment risks with adversarial perturbations.

\subsection{Risk Management}

Risk management is essential to protect investors in a volatile stock market. Over the decades, numerous studies have been conducted to analyze and mitigate the potential risks associated with stock investments~\cite{RM_01,RM_02}. One prominent approach is the application of modern portfolio theory (MPT)~\cite{MPT}, which emphasizes the importance of diversification in reducing overall portfolio risk. Furthermore, researchers have proposed various financial models such as the Capital Asset Pricing Model (CAPM)~\cite{CAPM} and Black-Scholes model~\cite{BS_Model} to better understand and quantify risk factors. Recently, advanced machine learning techniques have also found their significance on stock risk management~\cite{ML_risk_01} and some researchers have incorporated stock volatilities into reinforcement learning models to achieve better risk-adjusted profits~\cite{RL_01,RL_02}. However, most of the existing risk management methods are designed for some specific scenarios or models, which cannot be directly applied to the stock recommendation model. In this paper, we innovatively propose a risk management approach tailored for stock recommendation and demonstrate the feasibility of adversarial learning in risk management.

\subsection{Adversarial Training}

Desipite the excellent learning ability, deep neural networks (DNNs) have been found to be vulnerable to adversarial perturbations on data, i.e., adversarial examples~\cite{ADV_03,ADV_01}. Therefore, lots of adversarial training (AT) methods have been proposed to enhance the adversarial robustness of DNN classifiers~\cite{GAT,ADV_02}, such as FGSM~\cite{ADV_01}, PGD~\cite{PGD}, ADT~\cite{ADT}, etc. As for stock classification, Adv-ALSTM~\cite{adv_LSTM} might be the first work to engage adversarial training for stock movement prediction. All of the above conventional AT methods aim to improve model generalization by training the model to produce the \emph{same output} on both original and perturbed examples. In contrast, researchers of dialogue generation have proposed Inverse Adversarial Training (IAT) to encourage the model to be sensitive to perturbations and generate more diverse responses~\cite{IAT}. In this paper, we combine advantages of AT and IAT to better enhance the risk awareness of the stock recommendation model.

\subsection{Variational Autoencoder}

As one of the mainstream deep generative models, the variational autoencoder (VAE)~\cite{VAE} is good at learning the probability distribution of high-dimensional data through low-dimensional latent representations and generating high-quality data~\cite{VAE_survey01,VAE_survey02}. During the past few years, VAE has received widespread attention in various research fields and achieved promising results in image and audio generation~\cite{VAE_image01,VAE_image02,VAE_image03}, speech processing~\cite{VAE_speech01,VAE_speech02,VAE_speech03}, text processing~\cite{VAE_text} and biomedical informatics~\cite{VAE_bio,VAE_bio02}. In financial applications,~\cite{LSTM_VAE_stock} employs an LSTM-VAE framework to perform multi-step-ahead prediction of the stock closing price and~\cite{stocknet} develops a VAE-based model combining social media text and price signals for stock movement prediction. Recently, a more advanced model FactorVAE~\cite{FactorVAE} has been proposed to predict cross-sectional stock returns by regarding stock factors as the latent random variables in VAE. Inspired by previous work, we also employ the VAE architecture to generate representative risk indicators for our stock recommendation model.

\section{Conclusion}
\label{sec:conclusion}

In this paper, we propose a novel adversarial learning framework (SVAT) for risk-aware stock recommendation. In the first level, we design a split adversarial training method to enhance model's sensitivity to the adversarial perturbations of risky stock examples. In the second level, we devise a variational perturbation generator to model diverse risk factors and generate representative adversarial examples as risk indicators. Besides, the variational architecture enables our method to provide a rough risk quantification for investors, showing an additional advantage of interpretability. Experiments on three real-world datasets demonstrate that our method effectively reduces the volatility of the recommendation model and achieves the best risk-adjusted profit against 7 baselines. In addition, we demonstrate the efficiency of each component of the SVAT algorithm through ablative and qualitative experiments.

For the future research, we aim to explore the risk-aware adversarial learning for long-term stock prediction and incorporate additional data sources such as financial news for better risk modeling.

\begin{acks}
The research is supported by the National Natural Science Foundation of China (92370119, 62032025, 62376113), the Key-Area Research and Development Program of Shandong Province (2021CXGC010108), and Jiangsu Science and Technology Program (BE2020006-4).
\end{acks}

\bibliographystyle{ACM-Reference-Format}
\bibliography{TOIS-2023-0218}

\appendix

\end{document}